\documentclass[aps,pre,twocolumn,superscriptaddress,longbibliography,nobibnotes,nodoi,nourl,noeprint]{revtex4-2}
\usepackage{amsmath,amssymb,physics,bm,siunitx,booktabs}
\usepackage{xcolor,graphicx,soul,tikz}
\usepackage[colorlinks=true,citecolor=blue,urlcolor=blue,linkcolor=black]{hyperref}

\newcommand{\rms}{\rm\scriptscriptstyle}
\newcommand{\sss}{\scriptscriptstyle}
\newcommand{\kb}{k_{\rms B}}
\newcommand{\tepsA}{\tilde\epsilon^{\rm A}}
\newcommand{\tepsB}{\tilde\epsilon^{\rm B}}
\newcommand{\tepsm}{\tilde\epsilon^<}
\newcommand{\epsA}{\epsilon_{\rm A}}
\newcommand{\epsB}{\epsilon_{\rm B}}
\newcommand{\omA}{\omega_{\rm A}}
\newcommand{\omB}{\omega_{\rm B}}
\newcommand{\epsAp}{\epsilon_{\rm A}^\prime}
\newcommand{\epsBp}{\epsilon_{\rm B}^\prime}
\newcommand{\mAA}{\bar\epsilon^{\rm A}_{\rm A}}
\newcommand{\mAB}{\bar\epsilon^{\rm A}_{\rm B}}
\newcommand{\mBA}{\bar\epsilon^{\rm B}_{\rm A}}
\newcommand{\mBB}{\bar\epsilon^{\rm B}_{\rm B}}
\newcommand{\mybinom}[3][1.2]{\scalebox{#1}{$\binom{#2}{#3}$}}

\DeclareMathOperator*{\eqo}{\stackrel{\scriptscriptstyle\mathcal{O}}{=}}

\graphicspath{{figures/}}

\definecolor{darkred}{rgb}{0.7, 0, 0} 

\begin{document}

\title{Theory for the mixed alkali effect in glasses}

\author{Justus Leiber}
\email{jleiber@uni-osnabrueck.de}
\affiliation{Universit\"{a}t Osnabr\"uck, Institut f\"ur Physik, Barbarastra{\ss}e 7, D-49076 Osnabr\"uck, Germany}

\author{Quinn Emilia Fischer}
\email{qfischer@uni-osnabrueck.de}
\affiliation{Universit\"{a}t Osnabr\"uck, Institut f\"ur Physik, Barbarastra{\ss}e 7, D-49076 Osnabr\"uck, Germany}

\author{Sven Lohmann}
\email{slohmann@uni-osnabrueck.de}
\affiliation{Universit\"{a}t Osnabr\"uck, Institut f\"ur Physik, Barbarastra{\ss}e 7, D-49076 Osnabr\"uck, Germany}

\author{Philipp Maass}
\email[Corresponding author: ]{maass@uni-osnabrueck.de}
\affiliation{Universit\"{a}t Osnabr\"uck, Institut f\"ur Physik, Barbarastra{\ss}e 7, D-49076 Osnabr\"uck, Germany}

\date{May 13, 2026; revised July 10, 2026}

\begin{abstract}
The mixed alkali or mixed mobile ion effect in glasses manifests itself by 
strong nonlinear variations of ionic transport properties
upon mixing of different types of mobile ions. 
We develop a theory for this effect based on thermally
activated hopping transport in disordered site energy landscapes that
consistently incorporates the statistical‑mechanical and kinetic
aspects of a mobile ion mixture. This includes a consideration of the
joint probability density of site energy states, generalized Fermi
distributions for mean site occupations, and cross-terms in the
current response described by nondiagonal Onsager coefficients.  The
theory shows that a mixed alkali effect can arise even when the two
ion species share identical site energy distributions. It suffices that
sites have distinct energies when occupied by ions of different type.
Taking into account that a mismatch energy is needed for ions of one type to occupy sites adapted
to the other type, the mixed alkali effect becomes stronger.
Spatial correlations between site energies are needed for the mobility
of the majority ion to decrease stronger than exponential upon
replacement by the minority ion. The theory agrees well with kinetic
Monte Carlo simulations. Application to mixed alkali phosphate glasses
yields good agreement with measured conductivity activation energies.
\end{abstract}

\maketitle

\section{Introduction}

The mixed alkali effect (MAE) occurs in ion‑conducting
glasses when one type of mobile alkali ion is replaced by
another. Under such mixing, the mobility of the substituted ion
decreases, while that of the introduced ion increases
\cite{Isard:1969, Day:1976, Ingram:1987, Dyre/etal:2009}. Because the
effect manifests itself in the activation energies, the ion mobilities
at low temperature vary by several orders of magnitude as a function
of the mixing ratio $x$.

The MAE is a universal phenomenon. It has been observed in glasses
containing protons \cite{Molinelli/etal:1985, Garcia/etal:2002},
silver and thallium ions \cite{Bartholomew:1973, Carette/etal:1983,
  Machida/etal:1995, Rau/etal:2001, Bokova/etal:2015, Bokova/etal:2020},
and even in anion‑conducting glasses where fluorine ions are exchanged
by chlorine ions \cite{Almeida/Mackenzie:1982, Kahnt/Reau:1990,
  Alenezi/etal:2026}. Since the effect occurs for species other
than alkali ions, it is sometimes referred to as the mixed mobile ion
effect.  The activation energy of the total ionic conductivity
exhibits a maximum as a function of $x$. Other properties are likewise
affected by the MAE; for example, the internal friction measured in
mechanical‑relaxation experiments shows a pronounced dependence on the
mixing ratio \cite{Shelby:1965, Zdaniewski/etal:1979}.

Theories suggest that the mixed alkali effect arises because different
ion species interact differently with the surrounding host
network. Sites in the glass have distinct energies depending on which
ion type occupies them.  Building on this concept, the origin
of the effect can be understood in principle
\cite{Hunt:1999, Baranovskii/Cordes:1999, Maass:1999, Kirchheim:2000}.

The MAE was successfully modeled by molecular dynamics
simulations \cite{Habasaki/etal:1995, Habasaki/etal:2004,
  Balbuena/etal:2014, Lodesani/etal:2020, Noritake/Naito:2023}. In
particular, simulations of mixed alkali silica glasses
\cite{Lammert/Heuer:2005} have demonstrated the importance of distinct
energy landscapes for the MAE. NMR analyses of mixed‑ionic glasses
corroborate this finding \cite{Tsuchida/etal:2012,
  NogueiradosSantos/etal:2025}.

In this study, we further develop the theory of the MAE based on site
energy landscapes by consistently taking into account the statistical
mechanics and kinetics of the mixture of mobile ion types.  This
requires consideration of (i) the joint probability density of site
energy states, (ii) generalized Fermi distributions for the mean
occupation numbers that depend on the chemical potentials of both ion
species, and (iii) off‑diagonal Onsager kinetic coefficients
describing cross-terms in linear current–field relations.

When dealing with thermally activated ion transport in disordered site
energy landscapes, it is not permissible to treat the ions as
noninteracting, because at low temperatures they would all accumulate
on the lowest‑energy sites. In a minimal model, one therefore has to
introduce a site exclusion interaction, i.e., each site in the glass
network can be occupied by at most one ion. Our analysis relies on
such a minimal model, deliberately neglecting the Coulomb interaction
between the mobile ions. Due to its long-range nature, this Coulomb
contribution to the site energies is expected to vary far less than
the contribution arising from interactions with the network‑forming
ions. This assumption is supported by recent molecular dynamics
simulations \cite{Lohmann/etal:2026-tobepublished}.

For the minimal model, we generalize a theory for the many-particle
hopping processes in the presence of the site exclusion
interaction. Ionic conductivities obtained from kinetic Monte Carlo
simulations agree well with the predictions of this theory.

Remarkably, the MAE emerges even when the site energy distributions
for the two ion species are identical; it suffices that the two ion
types experience different energies on the same site. The effect becomes stronger,
when taking into account that an additional mismatch energy is needed for an ion of 
one type to occupy a site adapted to the other type.
This mismatch was viewed to be decisive for the occurrence of the MAE in early 
approaches based on hopping models \cite{Maass/etal:1992, Bunde/etal:1994}.
Because numerical calculations based on the theory can be performed in seconds
of computing time, it is possible to study systematically how the MAE
is affected by model parameters and to fit experimental data by
parameter optimization.

\section{Hopping transport in disordered site energy landscapes}
Ion transport in glasses proceeds by thermally activated
jumps of the mobile ions between sites, which are positions of local
minima in the disordered energy landscape. Energies of these minima vary
because the sites have different structural environments.

For modeling mixed ionic glasses, we consider $M$ sites, where
each site $i$ can be occupied by a at most one ion of type A or B, or by a
vacancy V. The fractions of occupied sites are
\begin{equation}
c_\alpha=\frac{N_\alpha}{M}\,,
\end{equation}
where $N_\alpha$ is
the number of each species, $\alpha\in\{{\rm A}, {\rm B},
{\rm V}\}$. It holds
$c_{\rm A}+c_{\rm B}+c_{\rm V}=1$.
The mixing ratio is
\begin{equation}
x=\frac{c_{\rm A}}{c_{\rm A}+c_{\rm B}}\,.
\end{equation}

We specify the occupation of sites by the numbers
\begin{equation}
n^\alpha_i=\left\{\begin{array}{cc}
1, & \mbox{if site $i$ is occupied by an $\alpha$ ion}, \\[0.5ex]
0, & \mbox{otherwise},
\end{array}\right.
\end{equation}
$n_i^{\rm A}+n_i^{\rm B}+n_i^{\rm V}=1$. Due to the site exclusion interaction,
it holds $n_i^\alpha n_i^\beta=n_i^\alpha\delta_{\alpha\beta}$.

The total energy of the system is
\begin{equation}
H(\mathbf{n})=H_0+\sum_{i=1}^M (\epsilon_i^{\rm A} n_i^{\rm A}+\epsilon_i^{\rm B} n_i^{\rm B})\,,
\label{eq:H2}
\end{equation}
where $H_0$ is the energy if all sites are vacant, and
$\epsilon_i^{\rm A}$ and $\epsilon_i^{\rm B}$ are the site energies of A and B ions.
The shortcut $\mathbf{n}$ specifies a microstate of the system,
$\mathbf{n}=((n_1^{\rm A},n_1^{\rm B}),(n_2^{\rm A},n_2^{\rm B}),\ldots,(n_M^{\rm A},n_M^{\rm B}))$.
For the hopping dynamics, only energy differences matter and the constant $H_0$ is irrelevant.

\begin{figure}[t!]
\centering
\includegraphics[width=0.75\columnwidth]{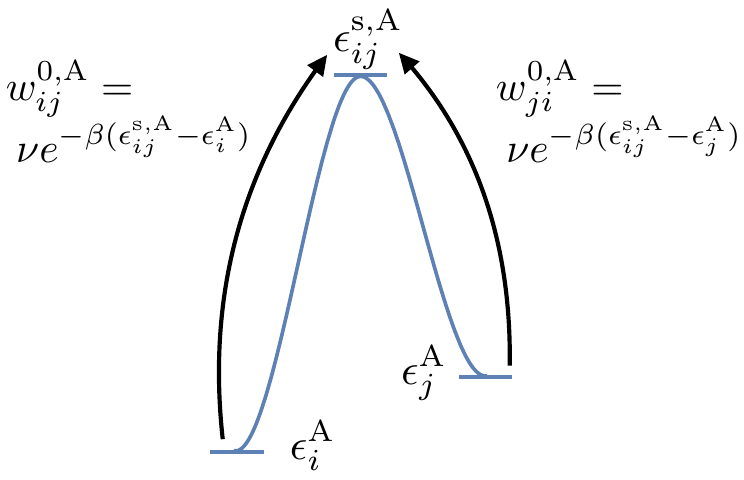}
\caption{Jump rates in the equilibrium state for an A ion between
nearest neighbor sites $i$ and $j$. The site energies are
$\epsilon_i^{\rm A}$, $\epsilon_j^{\rm A}$, and $\epsilon_{ij}^{\rm s, A}$
is the saddle point energy to be surmounted for jumps
between sites $i$ and $j$. For a jump from $i$ to $j$, the energy
barrier is $(\epsilon_{ij}^{\rm s, A}-\epsilon_i^{\rm A})$ and for
the jump from $j$ to $i$ it is $(\epsilon_{ij}^{\rm s, A}-\epsilon_j^{\rm A})$.
For jumps of a B ion between the same sites, the saddle point and 
site energies are different.}
\label{fig:illustration-jump-rates}
\end{figure}

A and B ions can jump to vacant nearest-neighbor sites. According to
chemical kinetics, the jump rates of an $\alpha$ ion occupying
site $i$ to a vacant nearest-neighbor site $j$ is
\begin{equation}
w_{ij}^{0,\alpha}=\nu_\alpha e^{-\beta(\epsilon_{ij}^{{\rm s},\alpha}-\epsilon^\alpha_i)}\,.
\label{eq:wij0}
\end{equation}
Here, $\nu_\alpha$ is an attempt frequency, $\beta=1/\kb T$ is the
inverse thermal energy, 
and $\epsilon_{ij}^{{\rm s},\alpha}=\epsilon_{ji}^{{\rm s},\alpha}$ is the saddle point
energy to be surmounted by an $\alpha$ ion in a thermally activated
transition between sites $i$ and $j$.  The rates are illustrated in
Fig.~\ref{fig:illustration-jump-rates} and satisfy the condition of
detailed balance,
\begin{equation}
e^{-\beta\epsilon^\alpha_i}w_{ij}^{0,\alpha}
=e^{-\beta\epsilon^\alpha_j}w_{ji}^{0,\alpha}\,.
\label{eq:detailed-balance-bare}
\end{equation}
This guarantees that in the stationary state of the hopping dynamics
with the rates in Eq.~\eqref{eq:wij0}, mircrostates $\mathbf{n}$ occur
with the equilibrium Boltzmann probability.

In the presence of a small external electric field ${\vb E}$, the jump
rates become
\begin{equation}
w_{ij}^{\alpha}=w_{ij}^{0,\alpha}\,\left(1+\frac{B_{ij}^\alpha}{2}\right)\,,
\label{eq:w}
\end{equation}
where
\begin{equation}
B_{ij}^\alpha=\beta q_\alpha\, \vb E\cdot \vb R_{ij}=-B_{ji}^\alpha\,.
\label{eq:B}
\end{equation}
Here, $q_\alpha$ is the charge of ions of type $\alpha$, and $\vb
R_{ij}=(\vb R_j\!-\!\vb R_i)$ is the vector pointing from position
$\vb R_i$ of site $i$ to position $\vb R_j$ of site $j$.

For calculating distributions of site energies, energy landscape
constructions can be applied that are based on knowledge of
concentrations and local charge distributions of glass forming units
\cite{Schuch/etal:2009, Schuch/etal:2011, Bosi/etal:2021, Bosi/Maass:2022}.

\section{Equilibrium properties}
\label{sec:equilibrium-properties}
The equilibrium distribution of microstates in the grand canonical ensemble is
\begin{equation}
p_{\rm eq}(\mathbf{n})=\frac{1}{\mathcal{Z}}
\exp[-\beta(H(\mathbf{n})-\mu_{\rm A}N_{\rm A}-\mu_{\rm B} N_{\rm B})]\,,
\label{eq:peq}
\end{equation}
where $\mathcal{Z}=\sum_{\mathbf{n}}\exp[-\beta H(\mathbf{n})]$ 
is the grand canonical partition sum, $\mu_\alpha$ is the chemical potential
of ions of type $\alpha\in\{{\rm A},{\rm B}\}$, and
\begin{equation}
\tilde\epsilon_i^\alpha=\epsilon_i^\alpha-\mu_\alpha\,.
\label{eq:tilde-eps-eq}
\end{equation}
Equilibrium averages with respect to $p_{\rm eq}(\mathbf{n})$ are
denoted as $\langle\ldots\rangle_{\rm eq}$ in the following.

As derived in Appendix~\ref{app:fermi-distributions}, the mean
occupation numbers of A ions in equilibrium are
\begin{equation}
\langle n_i^{\rm A}\rangle_{\rm eq}
=f(\beta\tilde\epsilon_i^{\rm A},\beta\tilde\epsilon_i^{\rm B})\,,
\label{eq:meanA-two-species}
\end{equation}
where
\begin{equation}
f(x,y)=\frac{1}{e^x+e^{x-y}+1}\,.
\label{eq:two-species-fermi-function}
\end{equation}
We call $f(x,y)$ the two-species Fermi function. It fulfills the
symmetry relation
\begin{equation}
e^x f(x,y)=e^y f(y,x)\,.
\label{eq_f-symmetry}
\end{equation}
The equation for $\langle n_i^{\rm B}\rangle_{\rm eq}$ follows by
interchanging A and B in Eq.~\eqref{eq:meanA-two-species},
\begin{equation}
\langle n_i^{\rm B}\rangle_{\rm eq}
=f(\beta\tilde\epsilon_i^{\rm B},\beta\tilde\epsilon_i^{\rm A})\,.
\label{eq:meanB-two-species}
\end{equation}
The mean occupation of sites by vacancies $V$ in thermal equilibrium
is
\begin{equation}
\langle n_i^{\rm V}\rangle_{\rm eq}
=1-\langle n_i^{\rm A}\rangle_{\rm eq}-\langle n_i^{\rm B}\rangle_{\rm eq}
=v(\beta\tilde\epsilon^{\rm A}_i,\beta\tilde\epsilon^{\rm B}_i)\,,
\label{eq:nV-v}
\end{equation}
where
\begin{equation}
v(x,y)=1-f(x,y)-f(y,x)=v(y,x)\,.
\label{eq:v-def}
\end{equation}
We refer to $v(x,y)$  as the vacancy Fermi function.
It holds
\begin{align}
v(x,y)=e^xf(x,y)=e^yf(y,x)=\frac{1}{1+e^{-x}+e^{-y}}\,.
\label{eq:v-f}
\end{align}
From Eqs.~\eqref{eq:meanA-two-species}, \eqref{eq:meanB-two-species},
\eqref{eq:v-def} and \eqref{eq:v-f} follows
\begin{equation}
\langle n_i^\alpha\rangle_{\rm eq}
=e^{-\beta\tilde\epsilon_i^\alpha}\langle n_i^{\rm V}\rangle_{\rm eq}\,,
\hspace{1em}\alpha={\rm A, B}\,.
\label{eq:nia-niV}
\end{equation}

For $\mu_{\rm B} \to -\infty$, the concentration of B ions becomes
zero, and we recover from Eqs.~\eqref{eq:meanA-two-species},
\eqref{eq:two-species-fermi-function} the standard Fermi distribution
for one component,
\begin{equation}
\langle n_i^{\rm A}\rangle_{\rm eq}
=\frac{1}{e^{\beta\tilde\epsilon^{\rm A}_i}+1}\,.
\end{equation}
The chemical potentials are fixed by the concentrations $c_\alpha$,
\begin{equation}
c_\alpha=\frac{1}{M}\sum_{i=1}^M \langle n_i^\alpha\rangle_{\rm eq}
=\frac{1}{M}\sum_{i=1}^M \sum_{\mathbf{n}} p_{\rm eq}(\mathbf{n})n_i^\alpha\,.
\label{eq:mu-determ-1}
\end{equation}
These are two coupled determining equations for 
$\mu_{\rm A}=\mu_{\rm A}(T,c_{\rm A},c_{\rm B})$ and
$\mu_{\rm B}=\mu_{\rm B}(T,c_{\rm A},c_{\rm B})$.

\begin{figure}[t!]
\includegraphics[width=0.72\columnwidth]{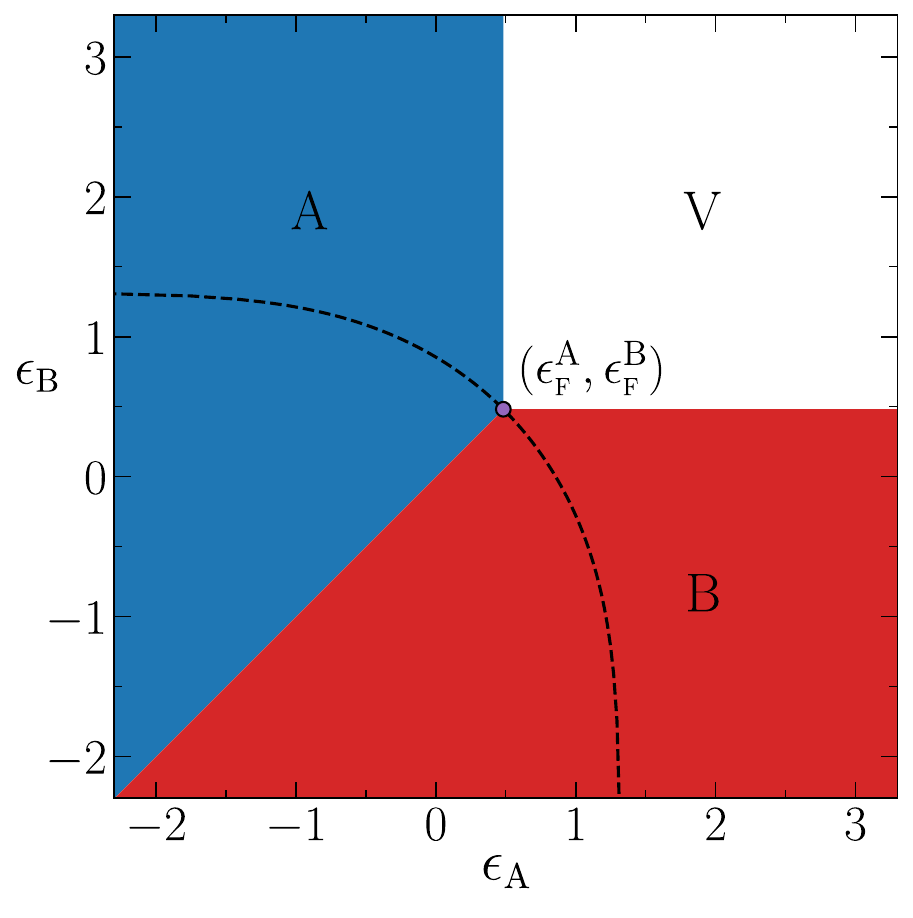}
\caption{Occupation of sites with energies $\epsilon_{\rm A}$ and
$\epsilon_{\rm B}$ in the zero-temperature limit. The blue, red and
white areas mark the pair of site energies $(\epsilon_{\rm A},\epsilon_{\rm B})$ 
occupied by A ions, B ions, and vacancies V, respectively.  The bullet marks 
the point $(\epsilon^{\rm A}_{\rms F},\epsilon^{\rm B}_{\rms F})$ of Fermi energies.
When the mixing ratio is changed between zero and one, it moves along a curve. The
dashed line gives an example of such a curve for the joint PDF of
site energies discussed in Sec.~\ref{sec:pdf-site-energies}
(parameters are the same as those in the uppermost, left panel of
Fig.~\ref{fig:joint_PDFs}).}
\label{fig:illustration-two-species-fermi-distribution}
\end{figure}

\subsection{Zero-temperature limit}
In the limit $T\to0$, the chemical potentials approach the Fermi
energies
\begin{equation}
\epsilon_{\rm F}^\alpha(c_{\rm A}, c_{\rm B})
=\lim_{T\to0} \mu_\alpha(T, c_{\rm A}, c_{\rm B})\,.
\end{equation}
For the mean occupation number $\langle n_i^{\rm A}\rangle_{\rm eq}$,
we obtain from Eqs.~\eqref{eq:tilde-eps-eq},
\eqref{eq:meanA-two-species} and \eqref{eq:two-species-fermi-function}
\begin{equation}
\lim\limits_{T\to0} \langle n_i^{\rm A}\rangle_{\rm eq}=\left\{\begin{array}{cc}
1\,, & \epsilon_i^{\rm A}<\epsilon_{\rms F}^{\rm A}\mbox{ and } 
(\epsilon_{\rms F}^{\rm A}-\epsilon_i^{\rm A})>(\epsilon_{\rms F}^{\rm B}-\epsilon_i^{\rm B})\,,\\[1ex]
0\,, &  \epsilon_i^{\rm A}>\epsilon^{\rm A}_{\rms F}\mbox{ or } 
(\epsilon_{\rms F}^{\rm A}-\epsilon_i^{\rm A})<(\epsilon_{\rms F}^{\rm B}-\epsilon_i^{\rm B})\,,
\end{array}\right.
\label{eq:nAeqTzero-1}
\end{equation}
and an analogous equation for $\lim_{T\to 0}\langle n_i^{\rm B}\rangle_{\rm eq}$ by 
interchanging A and B \footnote{When taking into consideration the marginal cases,
one obtains $\langle n_i^{\rm A}\rangle_{\rm eq}=1/2$ 
for  $\epsilon_i^{\rm A}=\epsilon^{\rm A}_{\rms F}$ and $\epsilon_i^{\rm B}>\epsilon^{\rm B}_{\rms F}$
as well as $\epsilon_i^{\rm A}<\epsilon^{\rm A}_{\rms F}$ and 
$\epsilon_i^{\rm A}-\epsilon_{\rms F}^{\rm A}=\epsilon_i^{\rm B}-\epsilon_{\rms F}^{\rm B}$, and
$\langle n_i^{\rm A}\rangle_{\rm eq}=1/3$ for $\epsilon_i^{\rm A}=\epsilon^{\rm A}_{\rms F}$ and 
$\epsilon_i^{\rm B}=\epsilon^{\rm B}_{\rms F}$.}.

Figure~\ref{fig:illustration-two-species-fermi-distribution} shows the
sites occupied by A ions, B ions, and vacancies $V$ in the
$\epsA,\epsB$-plane for $T=0$.  For finite 
$k_{\rms B}T<\min(\epsilon^{\rm A}_{\rms F},\epsilon^{\rm B}_{\rms F})$, the
distributions of mean occupation numbers are smeared out at the Fermi
edges $\epsilon_\alpha=\epsilon^\alpha_{\rms F}$ with a width of order
$k_{\rms B}T$.

\subsection{Joint probability density function\\ of site energies}
\label{subsec:joint-PDF}
In the thermodynamic limit, the determining equations
\eqref{eq:mu-determ-1} for the chemical potentials become
\begin{subequations}
\label{eq:mu-determ-2}
\begin{align}
\int\!\dd\epsilon_{\rm A}\! \int\!\dd\epsilon_{\rm B}\, 
\psi(\epsilon_{\rm A}, \epsilon_{\rm B}) f(\beta\tilde\epsilon_{\rm A}, \beta\tilde\epsilon_{\rm B})&=c_{\rm A}\,,
\label{eq:mu-determ-2-a}\\
\int\!\dd\epsilon_{\rm A}\! \int\!\dd\epsilon_{\rm B}\, 
\psi(\epsilon_{\rm A}, \epsilon_{\rm B}) f(\beta\tilde\epsilon_{\rm B}, \beta\tilde\epsilon_{\rm A})&= c_{\rm B}
\label{eq:mu-determ-2-b}\,.
\end{align}
\end{subequations}
Here, we have introduced the joint probability density function (PDF)
of states or site energies $\psi(\epsilon_{\rm A}, \epsilon_{\rm B})$:
$\psi(\epsilon_{\rm A}, \epsilon_{\rm B})\dd\epsilon_{\rm A}\dd\epsilon_{\rm B}$
is the probability of the A and B energies at
a site to lie in the intervals 
$[\epsilon_{\rm A}, \epsilon_{\rm A} +\dd\epsilon_{\rm A}]$ and 
$[\epsilon_{\rm B}, \epsilon_{\rm B} +\dd\epsilon_{\rm B}]$.

\subsection{Low-temperature behaviour\\ of chemical potentials}
\label{subseq:mu-for-low-T}
For reasoning the calculation of conductivity activation energies with
the extended AHL theory, discussed in
Sec.~\ref{subsec:generalized-AHL-theory} below, the low-$T$ behavior of
the chemical potentials $\mu_{\rm A}$ and $\mu_{\rm B}$ is important.
This behavior can be obtained by generalizing the Sommerfeld expansion
known for ideal Fermi gases with one component.  The derivation is given in
Appendix~\ref{app:low-T-behavior-mu}, and gives, as in the
one-component case, a leading term quadratic in $T$:
\begin{equation}
\mu_\alpha\sim \epsilon_{\rms F}^\alpha+\zeta_\alpha(k_{\rms B}T)^2\,,
\hspace{1em}\alpha\in\{{\rm A},{\rm B}\}\,.
\label{eq:mu-alpha-low-T}
\end{equation}
The $\zeta_\alpha$ are constants solely determined by the density of states. 

\section{Joint probability density function of site energies}
\label{sec:pdf-site-energies}
In previous studies of alkali borophosphate glasses, it was found that
changes of ionic transport properties can be captured by a
modeling with site energy distributions of nearly Gaussian form
\cite{Bosi/etal:2021, Bosi/Maass:2022}.  We therefore consider
Gaussian distributions for the PDF of site energies.  This choice is
also motivated by the central limit theorem when considering
that the energy of a site is formed additively from partial
contributions of its surroundings, where the contributions fluctuate
due to the structural disorder in the glassy network.

For mixed ionic glasses with two different types of mobile ions, we
need to specify the joint PDF introduced in
Sec.~\ref{subsec:joint-PDF}.  In the $\epsA,\epsB$-plane of site
energies in Fig.~\ref{fig:illustration-two-species-fermi-distribution},
energetically favorable ones for A (B) ions lie left (right) of the
diagonal $\epsA=\epsB$.  The numbers of sites favorable for A and B
ions should be proportional to the fractions $x$ and $(1\!-\!x)$ of A
and B ions. Accordingly, we set
\begin{equation}
\psi(\epsilon_{\rm A}, \epsilon_{\rm B})
=xG_{\rm A}(\epsilon_{\rm A}, \epsilon_{\rm B})
+(1-x)G_{\rm B}(\epsilon_{\rm A}, \epsilon_{\rm B})\,,
\label{eq:joint-PDF}
\end{equation}
where
\begin{align}
&G_\alpha(\epsA,\epsB)=\frac{1}{2\pi\sqrt{\det C^\alpha}}
\label{eq:Galpha}\\
&{}\times\exp[-\frac{1}{2}
\!\sum_{\beta,\gamma={\rm A,B}}\!(\epsilon_\beta\!-\!\bar\epsilon^\alpha_\beta) 
(C^\alpha)^{-1}_{\beta\gamma}(\epsilon_\gamma\!-\!\bar\epsilon^\alpha_\gamma)]\,,\hspace{0.5em}\alpha={\rm A,B}\,,
\nonumber
\end{align}
are bivariate Gaussian distributions with maxima in the energetically
favorable regions for A and B ions. This means that the pair of mean
values $(\mAA,\mAB)$ lies left of the diagonal $\epsA=\epsB$ in the
$\epsA,\epsB$-plane, and the pair of mean values $(\mBA,\mBB)$ right
of it. The covariance matrix $C^\alpha$ of the PDF
$G_\alpha(\epsA,\epsB)$ with elements
\begin{equation}
C^\alpha_{\beta\gamma}
=\langle(\epsilon_\beta-\bar\epsilon^\alpha_\beta)(\epsilon_\gamma-\bar\epsilon^\alpha_\gamma)\rangle
=C^\alpha_{\gamma\beta}
\end{equation}
can be written as
\begin{equation}
C^\alpha=
\begin{pmatrix}
C^\alpha_{\rm AA} & C^\alpha_{\rm AB}\\[0.5ex]
C^\alpha_{\rm AB} & C^\alpha_{\rm BB}
\end{pmatrix}
=
\begin{pmatrix}
\Delta_{\rm A}^{\alpha\,2} & \Delta_{\rm A}^\alpha\Delta_{\rm B}^\alpha\rho_\alpha\\[0.5ex]
\Delta_{\rm A}^\alpha\Delta_{\rm B}^\alpha\rho_\alpha & \Delta_{\rm B}^{\alpha\,2} 
\end{pmatrix}\,.
\end{equation}
Here, $\Delta_\beta^{\alpha\,2}$ is the variance of site energies
$\epsilon_\beta$, and $\rho_\alpha\in(-1,1)$ is the Pearson
correlation coefficient.

The parameters have the following physical significance: $\mAA$ and
$\mAB$ are the mean energies of A and B ions on sites with structural
environment favorably adapted to A ions. The difference
$\epsilon_{\rm mis}^{\rm B}=\mAB-\mAA\ge0$ hence is a mismatch energy quantifying
how much higher the energy of a B ion is on average, when it is on an
A adapted site. Likewise, $\mBB$ and $\mBA$ are the mean energies of B
and A ions on B adapted sites, with 
$\epsilon_{\rm mis}^{\rm A}=\mBA-\mBB\ge0$ the mismatch energy of an A ion.

The standard deviations $\Delta_{\rm A}^{\rm A}$ and $\Delta_{\rm B}^{\rm A}$
quantify how strongly A and B site energies on an A
adapted site fluctuate around their mean values, and 
$\Delta_{\rm B}^{\rm B}$ and $\Delta_{\rm A}^{\rm B}$ have the analogous meaning
for a B adapted site. The sign of $\rho_\alpha$ specifies whether
deviations of A and B energies from their mean values on an $\alpha$
adapted site are positively or negatively correlated, and the absolute
value of $\rho_\alpha$ signals how strong such correlations are.

For studying generic features of the model
in Secs.~\ref{subsec:evaluation-ahl}, \ref{subsec:activation-energy-landscape-features} below, 
we choose a symmetric parameter set with respect to A and B ions, i.e.\
$\mAA=\mBB$, $\mAB=\mBA$, $\Delta_{\rm A}^{\rm A}=\Delta_{\rm B}^{\rm B}$,
$\Delta_{\rm B}^{\rm A}=\Delta_{\rm A}^{\rm B}$, and $\rho_{\rm A}=\rho_{\rm B}$,
and further reduce the parameter number by setting
$\Delta_{\rm B}^{\rm A}=\Delta_{\rm A}^{\rm A}$. Without loss of generality,
we take $\mAA=0$ as zero point of energy
and $\Delta_{\rm A}^{\rm A}$ as energy unit,
\begin{equation}
\Delta_0\equiv\Delta_{\rm A}^{\rm A}\,.
\label{eq:Delta0}
\end{equation}
Only two parameters remain then, that are the mismatch energy
\begin{equation}
\epsilon_{\rm mis}\equiv\mAB-\mAA=\mBA-\mBB\ge0\,,
\label{eq:epsmis}
\end{equation}
and the onsite energy correlation
\begin{equation}
\rho\equiv\rho_{\rm A}=\rho_{\rm B}\in(-1,1)\,.
\label{eq:rho}
\end{equation}

\begin{figure}[t!]
\includegraphics[width=\columnwidth]{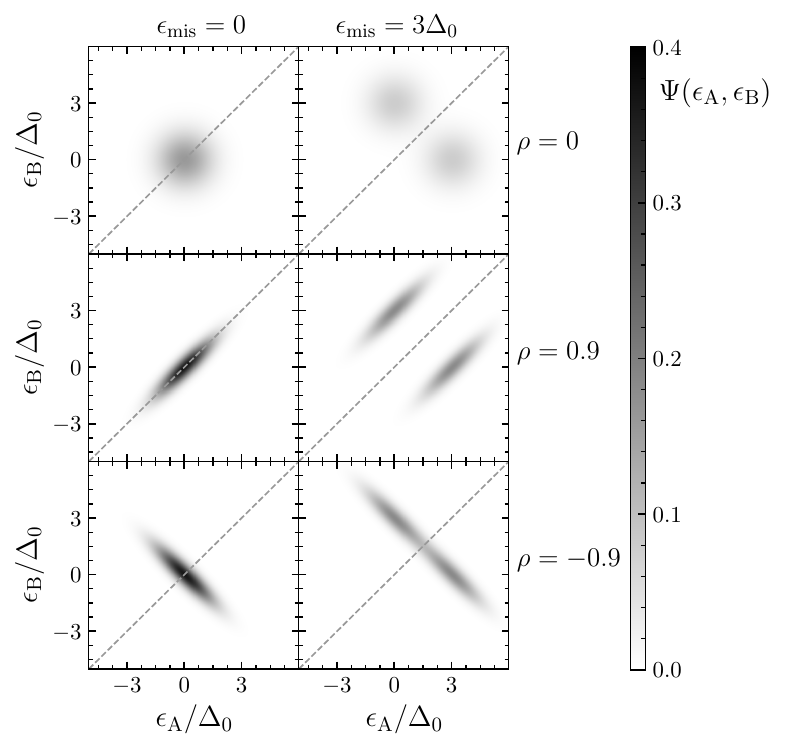}
\caption{\small Grey-scale plots of joint PDFs $\psi(\epsilon_{\rm A}, \epsilon_{\rm B})$ 
at a fixed mixing ratio $x=0.5$ for symmetric
parameters with respect to A and B ions. The left column shows PDFs
for vanishing mismatch energy $\epsilon_{\rm mis}=0$ and three
onsite energy correlations $\rho=0$ (no correlations), $\rho=0.9$
(strong positive correlations) and $\rho=-0.9$ (strong negative
correlations).  The right column shows PDFs for the same
$\rho$-values but mismatch energy $\epsilon_{\rm mis}=3$.  The
dashed lines in the $\epsilon_{\rm A}, \epsilon_{\rm B}$-planes
separate the areas energetically favorable for A ions (left of
diagonal) and B ions (right of diagonal).}
\label{fig:joint_PDFs}
\end{figure}

Figure~\ref{fig:joint_PDFs} shows joint PDFs of site energies at a
fixed mixing ratio $x=0.5$ for vanishing mismatch energy
$\epsilon_{\rm mis}=0$ (left column) and $\epsilon_{\rm mis}=3$
(right), and three values of onsite energy correlations $\rho=0$,
$\rho=0.9$ and $\rho=-0.9$. For vanishing $\epsilon_{\rm mis}$ and
$\rho$, the two bivariate Gaussians in Eq.~\eqref{eq:joint-PDF} are
equal and merge to one with circular isolines for all mixing ratio
$x$.  For $\epsilon_{\rm mis}>0$ and $\rho=0$, each of the
bivariate Gaussians in Eq.~\eqref{eq:joint-PDF} is circular, but their
maxima are separated with equal distance from the diagonal
$\epsilon_{\rm A}=\epsilon_{\rm B}$ for the symmetric parameter
set. With varying $x$,
the maxima of the two bivariate Gaussians change according to the 
weighting in Eq.~\eqref{eq:joint-PDF}.
For $\rho>0$ ($\rho<0$), the
circular isolines deform to elliptical ones with longer main axes
aligned in (perpendicular to) the direction of the diagonal. Positive
(negative) onsite energy correlations imply that deviations of
$\epsilon_i^{\rm A}$ from $\bar\epsilon^{\rm A}_{\rm A}$ or
$\bar\epsilon^{\rm B}_{\rm A}$ and deviations of $\epsilon_i^{\rm B}$
from $\bar\epsilon^{\rm A}_{\rm B}$ or $\bar\epsilon^{\rm B}_{\rm B}$
are likely of the same (different) sign. In the negatively correlated
case, a significant overlap of the two Gaussians can occur even for
strongly separated maxima due to a high $\epsilon_{\rm mis}$.

The symmetric setting with only two parameters allows us to
systematically investigate the influence of $\epsilon_{\rm mis}$ and
$\rho$ on ion transport properties.  For fitting experimental data,
asymmetry in the parameter sets needs to be taken into account.

When calculating chemical potentials in the limits $x\to0$ or $x\to1$,
one needs to take care that Gaussian distributions
are not bounded, i.e.\ the absolute values of site energies can be
arbitrarily large. For large positive site energies, this is no
problem because corresponding sites are not visited.
For large negative site energies, the
situation is different, because the Fermi energy 
$\epsilon_{\rm F}^\alpha$ of $\alpha$ ions approaches the minimal site energy
$\min\{\epsilon_i^\alpha\}$ when their concentration $c_\alpha$ goes
to zero, i.e.\ for either $x\to0$ or $x\to1$. In the thermodynamic
limit, this would cause the corresponding Fermi energy to become
negative infinite. Hence, when modeling transport properties of the
minority ion in the limit $x\to0$ or $x\to1$, one should take into
account that in reality site energies cannot become arbitrary
small. For the Gaussian distributions this is relevant only if the
concentration of the minority ion becomes extremely small.

\vspace*{-2ex}

\subsection{Spatial correlations between site energies}
\label{subsec:spatial-correlations}
Mixed ionic glasses often exhibit the so-called vulnerability effect
\cite{Ingram/Roling:2003, Moynihan/Lesikar:1981}.  This refers to a
stronger than exponential decrease of the total ionic conductivity
$\sigma(x)$ of glasses upon minor substitutions by a foreign ion.  The
strong decrease is reflected in concave curvatures of conductivity
activation energies for $x\to0$ or $x\to1$.

Short-range spatial correlations between site energies were suggested
to be responsible for the vulnerability
\cite{Peibst/etal:2005,Maass/Peibst:2006}.  They lead to a clustering
of sites adapted to the same type of ion. Upon initial replacement of
the majority by the minority ion, clusters of sites adapted to the
minority act as vacancy traps, reducing the mobility of the majority
ion strongly. The vacancy trapping occurs because in the clusters
adapted to the minority ion, it is energetically favorable to occupy
sites with a vacancy rather than a majority ion.

To generate spatially correlated site energies $\epsilon_i^{\prime\alpha}$,
we locally average over the spatially uncorrelated site energies $\epsilon_i^{\alpha}$,
\vspace{-3ex}
\begin{equation}
\epsilon_i^{\prime \alpha}
=(1-\kappa)\epsilon_i^\alpha + \kappa\frac{1}{z_i}\sum_{j\rm{NN}i} \epsilon_j^\alpha\,.
\label{eq:epsprime-eps}
\end{equation}
Here, $j{\rm NN}i$ denotes an average over the $z_i=\sum_{j\rm{NN}i}$
nearest neighbors of site $i$, and $\kappa\le 1$ is a
positive weighting coefficient controlling the strength of the
correlations.  The transformation \eqref{eq:epsprime-eps} leads to
modified PDFs $\psi'(\epsilon_i^{\prime \rm A}, \epsilon_i^{\prime \rm B})$,
which can be calculated from the PDFs of the uncorrelated site energies.

In the subsequent modeling we choose a simple cubic lattice for the
site positions.  While this is not realistic for a glass, the true
spatial arrangement of sites should be of minor importance. This is
because disorder in the glassy network is already accounted for by the
fluctuating site energies. In particular, long-range diffusion paths
of the mobile ions in the disordered site energy landscape have a
highly irregular geometry also for lattices, see also the discussion
after Eq.~\eqref{eq:u0ij} below.

For the simple cubic lattice with $z_i=6$, the joint PDF
$\psi'(\epsilon_i^{\prime \rm A}, \epsilon_i^{\prime \rm B})$ is
derived in Appendix~\ref{app:PDF-correlated-energies}.  We insert it
into the coupled equations~\eqref{eq:mu-determ-2-a} and
\eqref{eq:mu-determ-2-b} for determining the chemical potentials.

\section{Transport coefficients and generalized AHL theory}
\label{sec:transport-coefficients-AHL}

\subsection{Onsager coefficients and partial conductivities}
\label{subsec:Onsager-and sigmaalpha}
For analyzing transport of ions of different types A
and B, it is important to take into account that forces acting on
ions of type A not only generate currents of A ions but also
of B ions.  Such cross effects are captured by the nondiagonal
Onsager coefficients.

Let us imagine forces $f_\alpha$ acting solely on ions of type
$\alpha$ in the same direction.  In the linear response regime,
particle current densities $J_\alpha$ can then be written as
\begin{subequations}
\label{eq:currents-in-LIT}
\begin{align}
J_{\rm A}=J_{\rm A}(f_{\rm A},f_{\rm B})
=\frac{L_{\rm{AA}}}{T}f_{\rm A}+\frac{L_{\rm{AB}}}{T}f_{\rm B}\,,
\label{eq:currents-in-LIT-a}\\
J_{\rm B}=J_{\rm B}(f_{\rm A},f_{\rm B})
=\frac{L_{\rm{BA}}}{T}f_{\rm A}+\frac{L_{\rm{BB}}}{T}f_{\rm B}\,,
\label{eq:currents-in-LIT-b}
\end{align}
\end{subequations}
where $L_{\alpha\beta}$ are the Onsager coefficients. According to 
linear irreversible thermodynamics, they must satisfy
\begin{subequations}
\label{eq:onsager-relations}
\begin{align}
L_{\rm{AA}} \geq 0\,, \quad L_{\rm{BB}} \geq 0\,,\label{eq:onsager-relations-a}\\
L_{\rm{AB}}=L_{\rm{BA}}\leq \sqrt{L_{\rm{AA}}L_{\rm{BB}}}\,.
\label{eq:onsager-relations-b}
\end{align}
\end{subequations}

For ions with charge $q_\alpha=e$ driven by an electric field $E$, the forces
are equal, $f_{\rm A}=f_{\rm B}=eE$, and partial charge current
densities $eJ_\alpha$ satisfy
\begin{equation}
eJ_\alpha=\sigma_\alpha E\,,
\label{eq:sigmaalpha}
\end{equation}
where $\sigma_\alpha$ are the partial conductivities. Inserting $eE$
for the forces in Eq.~\eqref{eq:currents-in-LIT} and comparing with
Eq.~\eqref{eq:sigmaalpha} yields
\begin{equation}
\sigma_\alpha = \frac{e^2}{T}(L_{\alpha\rm A} + L_{\alpha\rm B})\,.
\label{eq:partial-conductivities}
\end{equation}
The total dc conductivity is the sum of the partial conductivities,
\begin{equation}
\sigma=\sigma_{\rm A} + \sigma_{\rm B}\,.
\label{eq:sigmatot}
\end{equation}

\subsection{Generalized AHL theory for\\ partial conductivities}
\label{subsec:generalized-AHL-theory}
Fermionic hopping models with site energy disorder were studied long
time ago for describing electron transport  in amorphous semiconductors. 
According to the theory by Miller and
Abrahams~\cite{Miller/Abrahams:1960}, dc-conductivities in these
models can be calculated approximately by a mapping to a random
resistor network. An alternative derivation of their result was given
later by Ambegaokar, Halperin and Langer
(AHL)~\cite{Ambegaokar/etal:1971}, who also discussed the calculation
of conductivity activation energies by percolation theory.  We here
generalize the AHL approach to the fermionic hopping model for
transport of different types of ions in glasses.

The directed link current $I_{i\to j}^\alpha$ of ion $\alpha$ from
site $i$ to a neighbor site $j$ is
\begin{equation}
I_{i\to j}^\alpha = w_{ij}^\alpha\langle n_i^\alpha n_j^{\rm V}\rangle\,,
\label{eq:Iij-directed}
\end{equation}
where $\langle.\rangle$ denotes an average in the nonequilibrium
steady state (NESS).  In a local equilibrium approximation, the
distribution of microstates $p_{\rm eq}(\mathbf n)$ in the NESS is
expressed by site-dependent local chemical potentials
$\mu_i^{\alpha}$. The differences between site energies
and chemical potentials in Eq.~\eqref{eq:tilde-eps-eq} are modified to
\begin{equation}
\hat\epsilon_i^\alpha = \epsilon_i^\alpha - \mu_i^\alpha\,.
\end{equation}
With these $\hat\epsilon_i^\alpha$ replacing the $\tilde\epsilon_i^\alpha$,
Eqs.~\eqref{eq:meanA-two-species}-\eqref{eq:v-f}
describe the
mean occupation numbers in the NESS.  Moreover, equal-time
correlations between occupation numbers in the NESS
factorize in this approximation,
\begin{equation}
\langle n_i^\alpha n_j^{\rm V}\rangle
\simeq \langle n_i^\alpha\rangle \langle n_j^{\rm V}\rangle\,,
\hspace{1em} i\neq j\,.
\end{equation}
Using this and Eq.~\eqref{eq:nia-niV}, we obtain in linear order of
$\delta w_{ij}^\alpha = w_{ij}^\alpha - w_{ij}^{0,\alpha}$,
$\delta\mu_i^\alpha = \mu_i^\alpha-\mu_\alpha$ and $\delta n_i^{\rm V}
= \langle n_i^{\rm V}\rangle -\langle n_i^{\rm V}\rangle_{\rm eq}$:
\begin{align}
I_{i\to j}^\alpha &\simeq 
w_{ij}^\alpha e^{-\beta\hat\epsilon_i^\alpha}\langle n_i^{\rm V}\rangle\langle n_j^{\rm V}\rangle\\
&\eqo I_{i\to j}^{0,\alpha}\left( 1 + \frac{\delta w_{ij}^\alpha}{w_{ij}^{0,\alpha}} + \beta\delta\mu_i^\alpha + \frac{\langle\delta n_i^{\rm V}\rangle}{\langle n_i^{\rm V}\rangle_{\rm eq}} + \frac{\langle\delta n_j^{\rm V}\rangle}{\langle n_j^{\rm V}\rangle_{\rm eq}} \right)\,.
\nonumber
\end{align}
Here, $I_{i\to j}^{0,\alpha} = w_{ij}^{0,\alpha} \langle
n_i^\alpha\rangle_{\rm eq} \langle n_i^{\rm V}\rangle_{\rm eq}$ is the
directed link current in the equilibrium state and the relation
``$\eqo$'' means ``equal in leading order with respect to expansion
variables'', i.e.\ here the biases $B_{ij}^\alpha$ from
Eq.~\eqref{eq:B}.  Since $I_{i\to j}^{0,\alpha}=I_{j\to i}^{0,\alpha}$
for the directed link currents in equilibrium, the net link current
$I_{ij}^\alpha$ of ion $\alpha$ between sites $i$ and $j$ is
\begin{equation}
I_{ij}^\alpha=I_{i\to j}^\alpha
-I_{j\to i}^\alpha\eqo I_{i\to j}^{0,\alpha}\left[\frac{\delta w_{ij}^\alpha}{w_{ij}^{0,\alpha}}
-\frac{\delta w_{ji}^\alpha}{w_{ji}^{0,\alpha}}
+\beta (\mu_i^\alpha - \mu_j^\alpha)\right].
\label{eq:Iijalpha}
\end{equation}
An alternative derivation of this equation based on the
fluctuation-dissipation theorem is given in
Appendix~\ref{app:FDT-derivation_Iij}.

Using Eq.~\eqref{eq:w}, we obtain
\begin{equation}
I_{ij}^\alpha\eqo 
\beta I_{i\to j}^{0,\alpha}\left[(\mu_i^\alpha-q_\alpha \mathbf{E}\cdot\mathbf{R}_i)
-(\mu_j^\alpha-q_\alpha \mathbf{E}\cdot\mathbf{R}_j)\right].
\label{eq:ec-potential-difference}
\end{equation}
This corresponds to a current resulting from a difference of electrochemical potentials
\begin{equation}
\mu_i^{\alpha, \rm{ec}}
=\mu_i^\alpha-q_\alpha \mathbf{E}\cdot\mathbf{R}_i\,.
\end{equation}
Rewriting Eq.~\eqref{eq:ec-potential-difference} in terms of a charge
current $q_\alpha I_{ij}^\alpha$ driven by the voltage $V_{ij}=
(\mu_i^{\alpha, \rm{ec}} - \mu_j^{\alpha, \rm{ec}})/q_\alpha$ yields
\begin{equation}
q_\alpha I_{ij}^\alpha = g_{ij}^\alpha V_{ij}^\alpha
\end{equation}
with conductances
\begin{equation}
g_{ij}^\alpha = \beta q_\alpha^2 I_{i\to j}^{0,\alpha}
=\beta q_\alpha^2 w_{ij}^{0,\alpha} \langle n_i^\alpha\rangle_{\rm eq} \langle n_j^{\rm V}\rangle_{\rm eq}\,.
\label{eq:gija}
\end{equation}
The partial hopping conductivities $\sigma_{\rm A}^{\rms AHL}$ and
$\sigma_{\rm B}^{\rms AHL}$ in the local equilibrium approximation
thus are equal to that of the conductivities of two distinct electric
networks with random conductances $g_{ij}^{\rm A}$ and $g_{ij}^{\rm
  B}$. For each ion type $\alpha$, the $g_{ij}^\alpha=g_{ji}^\alpha$
depend on the temperature and on both ionic concentrations $c_{\rm A}$
and $c_{\rm B}$ via the chemical potentials $\mu_\alpha$ determining
the $\langle n_i^\alpha\rangle$.  The link conductances exhibit short
range correlations, because $g_{ij}^\alpha$ and $g_{ik}^\alpha$ with
$j\ne k$ both depend on $\epsilon_i^\alpha$.

For the two random conductance networks, the conductivities
$\sigma_\alpha^{\rms AHL}$
can be determined by setting all sites at one boundary surface of the
sample at constant electric potential, all sites at the opposite
boundary surface at zero potential, and calculating the current
between the surfaces by solving Kirchhoff's equations.  The numerical
solution is efficiently carried out by iterative Kron
reduction. Details of this procedure for calculating conductivities
are given in Ref.~\cite{Bosi/Maass:2022}.

Generalizing our recent approach for calculating frequency-dependent
conductivities based on the AHL theory \cite{Lohmann/etal:2026}, the
many-particle hopping dynamics in the site energy landscape can
be mapped onto simpler hopping dynamics of independent particles in
an associated landscape with random, short-range correlated barriers.
In the barrier landscape, all sites have equal energy. To this end, we use
Eqs.~\eqref{eq:wij0} and \eqref{eq:nia-niV} to write the directed
link currents $I_{i\to j}^{0,\alpha}$ in Eq.~\eqref{eq:gija} as
\begin{equation}
I_{i\to j}^{0,\alpha}=
\nu_\alpha e^{-\beta(\epsilon_{ij}^{\rm s,\alpha}-\mu_\alpha)}
v(\beta\tepsA_i, \beta\tepsB_i) v(\beta\tepsA_j, \beta\tepsB_j)\,.
\end{equation}
This is also the rate for particles to pass the link $(ij)$ in either
direction in the equilibrium state and can be identified with the
hopping rate $\nu_\alpha\exp(-\beta u_{ij}^\alpha)$ for surmounting a
barrier $u_{ij}^\alpha$, yielding
\begin{equation}
u_{ij}^\alpha = \epsilon_{ij}^{\rm s,\alpha}
- \mu_\alpha - \frac{1}{\beta}\ln\left[v(\beta\tepsA_i, \beta\tepsB_i)\right]
- \frac{1}{\beta}\ln\left[v(\beta\tepsA_j, \beta\tepsB_j)\right].
\label{eq:uija}
\end{equation}
Like the conductances in Eq.~\eqref{eq:gija}, the
$u_{ij}^\alpha=u_{ji}^\alpha$ depend on temperature and both ion
concentrations and exhibit short-range correlations.

Analysis of the hopping dynamics in the associated barrier landscape
allows one to go beyond dc transport and to determine
frequency-dependent conductivities for mixed ionic glasses.  There
exists an efficient numerical velocity autocorrelation (VAC) method
for calculating them \cite{Schroeder:2008}. In this method,
the problem is reduced to solving an inhomogeneous system of
linear equations.  An analysis of ac conductivities will be devoted to
later studies.

Here, we focus on dc transport and present results for the partial conductivities obtained
from the VAC method in the low-frequency limit. We have checked that these results agree with
the ones obtained from the alternative solution of Kirchhoff's equations described above.

Activation energies $E_\alpha^{\rms AHL}$ of the partial
conductivities $\sigma_\alpha^{\rms AHL}$ can be calculated by
considering the $T\to0$ limit of the link conductances $g_{ij}^\alpha$
from Eq.~\eqref{eq:gija} and applying percolation theory.

According to Eq.~\eqref{eq:mu-alpha-low-T}, the chemical potentials
$\mu_\alpha$ vary quadratically with $T$ for low temperatures, which
implies that the we can replace the $\mu_\alpha$ by the Fermi energies
$\epsilon_{\rms F}^\alpha$ in the variables
$\beta\tilde\epsilon_i^\alpha$ for $T\to0$ ($\beta\to\infty$).  In the
subsequent equations \eqref{eq:vfermi-lowT}-\eqref{eq:uij0alpha}, we
understand $\tilde\epsilon^\alpha_i$ as
$\tilde\epsilon^\alpha_i=\epsilon_i^\alpha-\epsilon_{\rm F}^\alpha$.
The asymptotic low-temperature behavior of the vacancy Fermi function
is
\begin{align}
&v(\beta\tepsA_i, \beta\tepsB_i)
=\frac{1}{1+e^{-\beta\tepsA_i}+e^{-\beta\tepsB_i}}
\label{eq:vfermi-lowT}\\
&\hspace{3em}\sim
\left\{\begin{array}{cc}
e^{\beta\min(\tepsA_i,0)}\,, & \tepsA_i<\tepsB_i\\[1ex]
e^{\beta\min(\tepsB_i,0)}\,, & \tepsA_i>\tepsB_i
\end{array}\right\}=e^{\beta\min[\tepsm_i,0]}\,,
\nonumber
\end{align}
where
\begin{equation}
\tepsm_i=\min(\tepsA_i,\tepsB_i)\,.
\label{eq:tepsm}
\end{equation}
Inserting this into Eq.~\eqref{eq:uija} and using $\langle
n_i^\alpha\rangle_{\rm eq}=e^{-\beta\tilde\epsilon_i^\alpha}\langle
n_i^{\rm V}\rangle_{\rm eq}$ from Eq.~\eqref{eq:nia-niV}, we obtain
for the barriers $u_{0,ij}^\alpha$ in the $T\to0$ limit
\begin{equation}
u_{0,ij}^\alpha=\epsilon_{ij}^{{\rm s}, \alpha}
-\epsilon_{\rm F}^\alpha-\min[\tepsm_i,0]-\min[\tepsm_j,0]\,.
\label{eq:uij0alpha}
\end{equation}

The activation energies $E_\alpha^{\rm AHL}$ then follow from the
critical thresholds for links with $u_{ij}^\alpha\le u$ to percolate:
\begin{equation}
E_\alpha^{\rms AHL}=\min_u\left\{u \,\Bigl|\, \parbox[c]{0.52\columnwidth}
{set of links $(ij)$ with $u_{0,ij}^\alpha\le u$\\ forms a percolating path}\right\}\,.
\label{eq:Ealpha}
\end{equation}
We have determined clusters of linked sites with $u_{ij}^\alpha\le u$
and the percolation thresholds $E_\alpha^{\rms AHL}$ by applying the
Hoshen-Kopelman algorithm \cite{Hoshen/Kopelman:1976}.

\section{Kinetic Monte Carlo simulations}
\label{sec:KMC-simulations}
For obtaining the Onsager coefficients in
Eq.~\eqref{eq:currents-in-LIT}, we perform continuous-time kinetic
Monte Carlo (KMC) simulations of the hopping dynamics by employing the
first-reaction time algorithm \cite{Gillespie:1978,
  Holubec/etal:2011}.  These simulations are carried out for a simple
cubic lattice with lattice constant $a$ and $(L/a)^3$ sites. Periodic
boundary conditions are applied in all directions. The sites are
occupied with $N_\alpha=c_\alpha L^3$ ions of type $\alpha$.  System
sizes lie in the range $L/a=30$-50. We checked that the results are
not affected by finite size effects.

For generating the energy landscapes according to the joint PDF
$\psi(\epsilon_{\rm A},\epsilon_{\rm B})$ in Eq.~\eqref{eq:joint-PDF},
random site energies $(\epsilon_i^{\rm A},\epsilon_i^{\rm B})$ are
assigned to each site $i$. This is done by choosing site $i$ to be
A-like with probability $x$ and B-like with probability $(1-x)$, and
then drawing $(\epsilon_i^{\rm A},\epsilon_i^{\rm B})$ from
$G_\alpha(\epsilon_{\rm A},\epsilon_{\rm B})$ in Eq.~\eqref{eq:Galpha}
for an $\alpha$-like site. When considering spatially correlated
landscapes, we in addition apply the transformation in
Eq.~\eqref{eq:epsprime-eps}.

\begin{figure}[t!]
\centering
\includegraphics[width=\columnwidth]{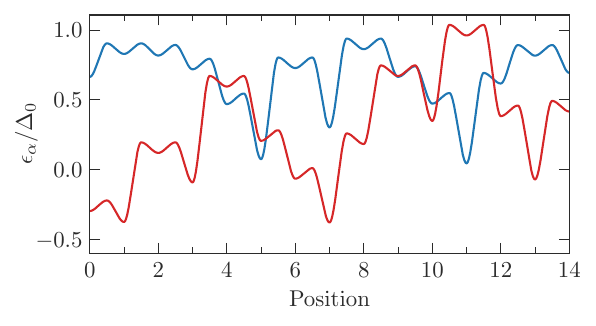}
\caption{One-dimensional sketch of
energy landscapes for A (blue line) and B (red line) ions in the constant lower barrier model for a
mixing ratio $x=0.2$.  The parameters are the ones used for
fitting the activation energies of mixed rubidium-lithium phosphate glasses 
Rb$_x$-Li$_{(1\!-\!x)}$-PO$_3$ in Sec.~\ref{subsec:modeling-experimental-observations},
see Table~\ref{tab:fit_parameters}.}
\label{fig:illustration-CLB-landscape}
\end{figure}

For the jump rates $w_{ij}^{0,\alpha}$ in Eq.~\eqref{eq:wij0}, we need
to specify the saddle point energies $\epsilon_{ij}^{{\rm
    s},\alpha}$. As these must be larger than both site energies
$\varepsilon_i^\alpha$ and $\varepsilon_j^\alpha$, we write
\begin{equation}
\varepsilon_{ij}^{\rm s, \alpha}
=\max(\varepsilon_i^\alpha,\varepsilon_j^\alpha)+u_{ij}^\alpha\,,
\label{eq:saddles-clb}
\end{equation}
where $u_{ij}^\alpha>0$ is the lower barrier for a jump of an $\alpha$ 
ion between sites $i$ and $j$. These lower barriers
have some distribution and it is possible that they
are correlated with the site energies.
For simplicity, we consider uniform barriers $u_{ij}^\alpha=u_0$ here,
equal for both types of ions. A one-dimensional sketch of an energy
landscape for this constant lower barrier model is shown
in Fig.~\ref{fig:illustration-CLB-landscape}. It exhibits both varying 
energy barriers and site energies, thus capturing essential features of 
complex landscape models.

Inserting Eq.~\eqref{eq:saddles-clb} into Eq.~\eqref{eq:wij0}, we
obtain
\begin{equation}
w_{ij}^{0,\alpha}=\nu'_\alpha\min(1,e^{-\beta(\varepsilon_j^\alpha-\varepsilon_i^\alpha)})\,,
\label{eq:metropolis-rates}
\end{equation}
where $\nu'_\alpha=\nu_\alpha e^{-\beta u_0^\alpha}$.
These are jump rates of the widely used
Metropolis form with an effective attempt frequency $\nu'_\alpha$.
The modified jump rates $w_{ij}^{\alpha}$
in the presence of an electric field follow from Eqs.~\eqref{eq:w}, \eqref{eq:B}.

Regarding the size of $u_0$, one can estimate that it should be small,
in particular much smaller than the activation energies of partial
conductivities. This follows from the fact that experimental
conductivity spectra for high frequencies $\omega\gtrsim1\si{GHz}$
tend to merge, a feature that is reproduced for a constant attempt
frequency. In the following, we set $u_0=0$,
i.e.\ $\nu'_\alpha=\nu_\alpha$. This setting is not crucial, because
activation energies of the partial conductivities would otherwise be
shifted by a small additive constant $u_0>0$. For simplicity, we
furthermore take the attempt frequencies for A and B ions equal,
\begin{equation}
\nu\equiv\nu_{\rms A}=\nu_{\rms B},
\label{eq:nu}
\end{equation}
and choose $\nu^{-1}$ as the time unit.

Using
\begin{subequations}
\begin{gather}
\min(x,y)=\frac{1}{2}(x+y-|x-y|)\,,\\
\max(x,y)=-\min(-x,-y)=\frac{1}{2}(x+y+|x-y|)\,,
\end{gather}
\end{subequations}
the zero-temperature link barriers $u_{0,ij}^\alpha$ [Eq.~\eqref{eq:uij0alpha}]
associated with the constant lower barrier model can be expressed as
\begin{align}
u_{0,ij}^\alpha=&\frac{1}{2}\left(|\epsilon_i^\alpha-\epsilon_j^\alpha|
+\left|\epsilon_i^\alpha-\max(\epsilon_{\rms F}^{\rm A},\epsilon_{\rms F}^{\rm B})\right|\right.
\label{eq:u0ij}\\
&
\left.\hspace{1em}
{}+\left|\epsilon_j^\alpha-\max(\epsilon_{\rms F}^{\rm A},\epsilon_{\rms F}^{\rm B})\right|
\right)
+\max(0,\epsilon_{\rms F}^\beta-\epsilon_{\rms F}^\alpha)\,.
\nonumber
\end{align}
This formula shows that the link barriers $u_{0,ij}^\alpha$ have a
global, link-independent additive contribution $\max(0,\epsilon_{\rms
  F}^\beta-\epsilon_{\rms F}^\alpha)$ for the ion with smaller Fermi
energy and that these barriers are low if both site energies
$\epsilon_i^\alpha$ and $\epsilon_j^\alpha$ are close to the Fermi
energy.  Accordingly, the dynamics of the mobile ions at low $T$ is
taking place on a subset of sites with site energies close to the
larger of the two Fermi energies. This subset has a strongly
irregular, fractal-like geometry, implying that the regular lattice
structure chosen in the modeling should not matter for long-range
transport properties.

To calculate the Onsager coefficients, we apply small forces $f_{\rm
  A}$ and $f_{\rm B}$ in the $x$-direction solely acting on the A and B
ions.  The jump rates $w_{i,j}^\alpha$ under these forcings are given
by Eq.~\eqref{eq:w} with $B_{ij}^\alpha=\beta f_\alpha a
b_{ij}^\alpha$ from Eq.~\eqref{eq:B}, where $b_{ij}^\alpha=\pm 1$ for
jumps in the $\pm x$-direction, and $b_{ij}^\alpha=0$ for jumps in the 
$y$- or $z$-direction.

After the system has evolved into a steady state, we determine current
densities by counting the numbers $N_\alpha^\pm$ of jumps by $\alpha$
ions in the $\pm x$-direction over a long time interval $\Delta t$.  The
net numbers of $\alpha$ ions passing one lattice plane perpendicular
to the forces per unit time are $(a/L)(N_\alpha^+ - N_\alpha^-)/\Delta
t$, yielding particle current densities
\begin{equation}
J_\alpha = \frac{a}{L^3}\frac{N_\alpha^+ -N_\alpha^-}{\Delta t}\,.
\end{equation}

For $f_\alpha a\lesssim0.4k_{\rms B}T$ we find the response to be in
the linear response regime. Using Eqs.~\eqref{eq:currents-in-LIT-a},
\eqref{eq:currents-in-LIT-b}, the Onsager coefficients follow from
\begin{subequations}
\begin{align}
\frac{L_{\rm AA}}{T}=\frac{J_{\rm A}(f_{\rm A},0)}{f_{\rm A}}\,,\hspace{1em}
\frac{L_{\rm BA}}{T}=\frac{J_{\rm B}(f_{\rm A},0)}{f_{\rm A}}\,,\\
\frac{L_{\rm AB}}{T}=\frac{J_{\rm A}(0,f_{\rm B})}{f_{\rm B}}\,,\hspace{1em}
\frac{L_{\rm BB}}{T}=\frac{J_{\rm B}(0,f_{\rm B})}{f_{\rm B}}\,.
\end{align}
\end{subequations}
In all cases we verified the relations \eqref{eq:onsager-relations-a} and \eqref{eq:onsager-relations-b}.

An alternative way to determine the Onsager coefficients is to use the
Einstein fluctuation formulas \cite{Allnatt/Lidiard:1993}
\begin{equation}
L_{\alpha\beta}=\frac{1}{6L^3k_{\rms B}T}\lim\limits_{t\to\infty}\sum_{i,j}
\frac{\langle \Delta\vb{r}_i^\alpha(t)\cdot  \Delta\vb{r}_j^\beta(t)\rangle_{\rm eq}}{t}\,,
\label{eq:Lab-fluctuation-formula}
\end{equation}
where $\Delta\vb r_i^\alpha(t)$ is the displacement of the $\alpha$
ion $i$ in time $t$.  Performing simulations in the equilibrium state
with the jump rates $w_{ij}^{0,\alpha}$ and determining the
displacements $\Delta\vb{r}_i^\alpha$ in a long time interval $t$,
yield the $L_{\alpha\beta}$ in Eq.~\eqref{eq:Lab-fluctuation-formula}.
Results of corresponding calculations agree with the forcing method.
We found the forcing method to be more efficient for our problem.

\begin{figure*}[t!]
\centering
\includegraphics[width=\linewidth]{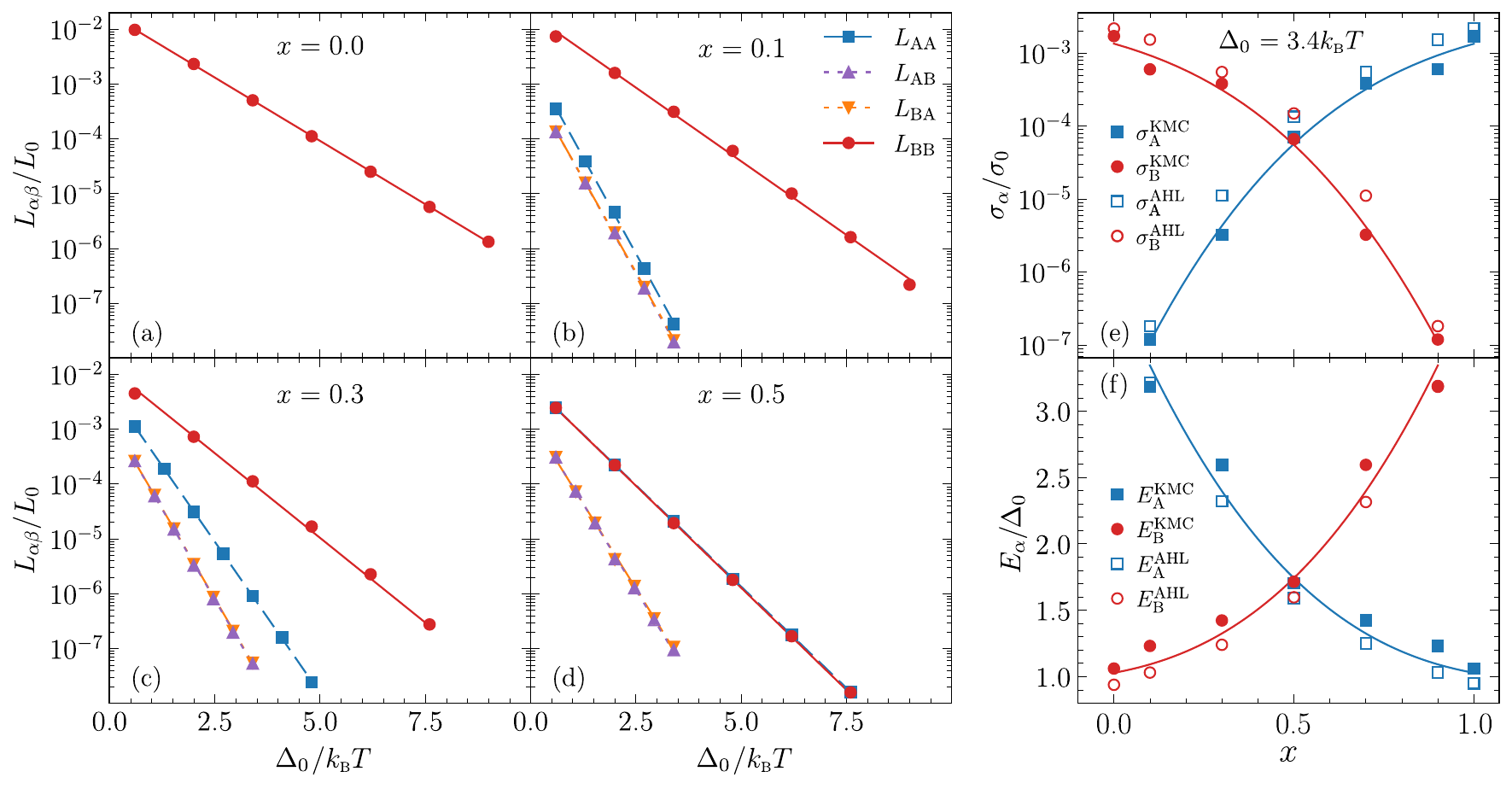}
\caption{\small Comparison of results from KMC simulations with the generalized AHL theory for the
site energy PDF with symmetric parameter setting and parameter values
$\epsilon_{\rm mis}/\Delta_0=3$ and $\rho=0$. (a)-(d)~Arrhenius plots of Onsager
coefficients $L_{\alpha\beta}$ determined from KMC simulations for four mixing ratios $x$.
(e) Isotherms of KMC simulated partial conductivities 
$\sigma_{\rm A}^{\rms KMC}(x)$, $\sigma_{\rm B}^{\rms KMC}(x)$,
and $\sigma_{\rm A}^{\rms AHL}(x)$, $\sigma_{\rm B}^{\rms AHL}(x)$
calculated from the generalized AHL theory for $\kb T=\Delta_0/3.4$. 
(f) Simulated activation energies $E_{\rm A}^{\rms KMC}(x)$, $E_{\rm B}^{\rms KMC}(x)$
and calculated $E_{\rm A}^{\rms AHL}(x)$, $E_{\rm B}^{\rms AHL}(x)$.}
\label{fig:RSEM_vs_ARBM}
\end{figure*}

\section{Results}
\label{sec:results}
We first show in Sec.~\ref{subsec:evaluation-ahl} that partial
conductivities $\sigma_\alpha^{\rms AHL}$ predicted by the generalized
AHL theory agree well with those calculated from KMC simulations.
Based on these findings, we analyze in
Sec.~\ref{subsec:activation-energy-landscape-features} how the
variation of conductivities and their activation energies with the
mixing ratio depends on the mismatch energy $\epsilon_{\rms mis}$
[Eq.~\eqref{eq:epsmis}], the onsite energy correlation $\rho$
[Eq.~\eqref{eq:rho}], and spatial correlations of the site energies
controlled by $\kappa$ in Eq.~\eqref{eq:epsprime-eps}.  This knowledge
is then used to find suitable parameters to fit experimental results
for various mixed alkali glasses in
Sec.~\ref{subsec:modeling-experimental-observations}.

Results for Onsager coefficients and conductivities are given in units of
\begin{align}
L_0=\frac{\nu}{\kb a}\,,\\
\sigma_0=\frac{\nu e^2}{a\Delta_0}\,,
\end{align}
where the lattice constant $a$ plays the role of the mean jump
distance, $\nu$ is the attempt frequency from Eqs.~\eqref{eq:wij0},
\eqref{eq:nu}, and $\Delta_0$ is the energy unit introduced in
Eq.~\eqref{eq:Delta0}.

For evaluating the generalized AHL theory in
Sec.~\ref{subsec:evaluation-ahl} and analyzing the relation between
energy landscape features and behavior of activation energies in
Sec.~\ref{subsec:activation-energy-landscape-features}, we choose the
symmetric parameter setting from Sec.~\ref{sec:pdf-site-energies}
with respect to A and B ions, including
the possibility of spatial correlations [$\kappa>0$ in
  Eq.~\eqref{eq:epsprime-eps}]. Due to the symmetry, we can restrict
the analysis to the range $0\le x\le 0.5$ of mixing ratios, where the
B ions are in majority.

\subsection{Evaluation of generalized AHL theory}
\label{subsec:evaluation-ahl}
Representative Arrhenius plots of simulated Onsager coefficients for
the energy landscape from Fig.~\ref{fig:joint_PDFs}(c) [$\epsilon_{\rm
    mis}=3$, $\rho=0$, $\kappa=0$] are shown in
Figs.~\ref{fig:RSEM_vs_ARBM}(a)-(c). As required, $L_{\rm AB}=L_{\rm
  BA}$, and it holds $L_{\rm AB}\ll\sqrt{L_{\rm AA}L_{\rm BB}}$. In
all cases, the $L_{\alpha\beta}$ follow an Arrhenius law.  The
activation energy $E_{\rm B}$ of $L_{\rm BB}$ (majority ion) is
smaller than $E_{\rm A}$ of $L_{\rm AA}$ (minority ion). With rising
substitution of the B by A ions, $E_{\rm B}$ increases and $E_{\rm A}$
decreases.  The activation energy $E_{\rm AB}$ of the nondiagonal
Onsager coefficient $L_{\rm AB}$ is larger than that of the diagonal
ones for all $x$.

Because $E_{\rm AB}>E_{\alpha}$, Eq.~\eqref{eq:partial-conductivities}
implies that the activation energies $E_\alpha$ of the
$L_{\alpha\alpha}$ are also those of the partial conductivities
$\sigma_\alpha$. Figure~\ref{fig:RSEM_vs_ARBM}(d) shows that they are
well predicted by the generalized AHL theory.

\begin{figure*}[t!]
\centering
\includegraphics[width=\linewidth]{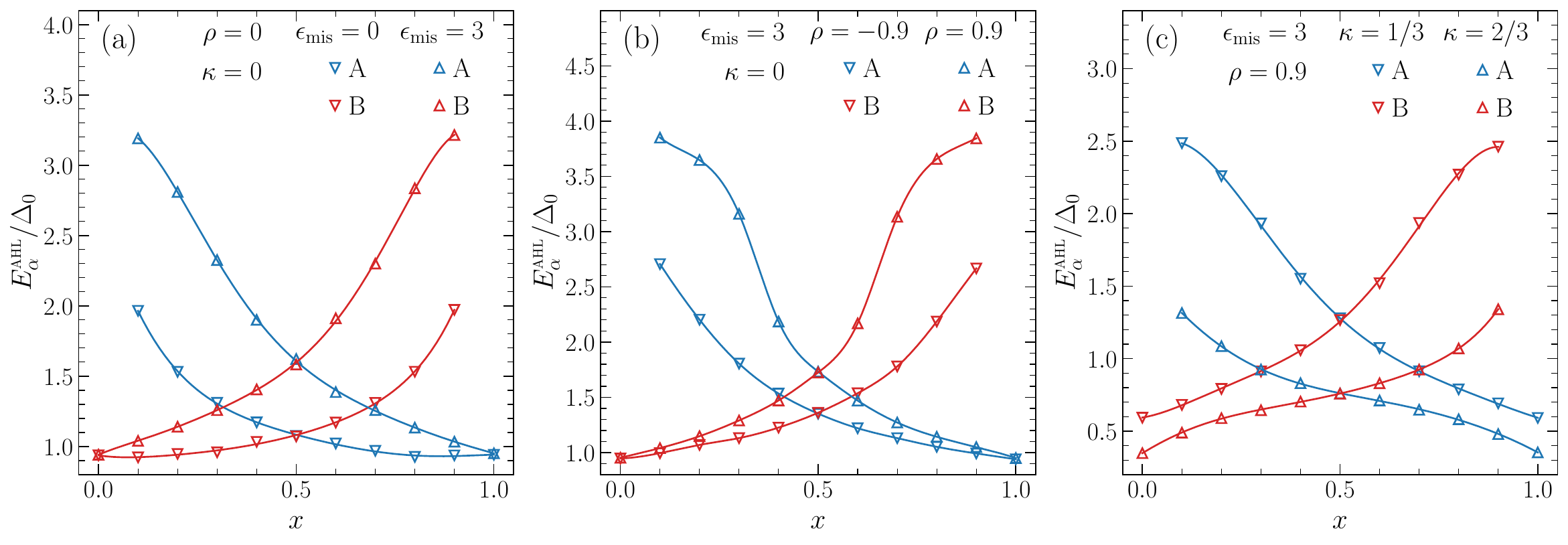}
\caption{\small Impact of energy landscape features on the variation
of activation energies $E_{\rm A}^{\rms AHL}(x)$ and
$E_{\rm B}^{\rms AHL}(x)$ of partial conductivities: data for (a) two mismatch energies
$\epsilon_{\rm mis}$ [Eq.~\eqref{eq:epsmis}], $\rho=0$ [no onsite
energy correlations, Eq.~\eqref{eq:rho}] and $\kappa=0$ [no
spatial correlations between site energies, Eq.~\eqref{eq:epsprime-eps}],
(b) $\epsilon_{\rm mis}=3$, $\kappa=0$, and two onsite energy correlations
$\rho=-0.9$ and $\rho=0.9$, and (c) $\epsilon_{\rm mis}=3$, $\rho=0.9$,
and two strengths $\kappa=1/3$ and $\kappa=2/3$ of spatial correlations
between site energies. Lines are guides to the eye.}
\label{fig:model_parameters}
\end{figure*}

We furthermore compare partial conductivities $\sigma_\alpha$,
obtained from the simulated $L_{\alpha\beta}$ by
Eq.~\eqref{eq:partial-conductivities}, with the partial conductivities
$\sigma_\alpha^{\rms AHL}$.  Isotherms of partial conductivities in
Fig.~\ref{fig:RSEM_vs_ARBM}(e) are well predicted by the generalized
AHL theory.
Simulated data are only slightly overestimated.
The more mobile ion governs the total ionic conductivity
$\sigma$ in Eq.~\eqref{eq:sigmatot}, which has the activation energy
\begin{equation}
E_\sigma=\min(E_{\rm A},E_{\rm B})\,.
\label{eq:Esigma}
\end{equation}
We thus can study the transport properties based on the
generalized AHL theory without performing extensive KMC simulations. 

\subsection{Impact of energy landscape features on the variation of activation energies}
\label{subsec:activation-energy-landscape-features}
An important question is which salient features of the energy
landscape determine the behavior of the activation energies
$E_\alpha(x)$ with the mixing ratio $x$. While the rise of activation
energy $E_\alpha$ of one type of ion upon substitution by the other
type is universal, the change of rise, reflected in the second
derivatives $E_\alpha''(x)=\dd^2 E_\alpha(x)/\dd x^2$, can be
different. Experimental data for activation energies of individual ion
mobilities are rare, because tracer diffusion measurements, suitable
for accessing such energies, are laborious.

More studied is the activation energy $E_\sigma(x)$ of the total ionic
conductivity $\sigma(x)$.  As discussed in
Sec.~\ref{subsec:spatial-correlations}, this often reflects the
vulnerability effect, i.e.\ concave curvatures of $E_\sigma(x)$ at
small $x$ or $(1-x)$.  Typically, a concave curvature of the
conductivity activation energy $E_\sigma(x)$ and a convex curvature of
$\ln\sigma(x)$ is observed over the entire range of mixing ratios
$x\in(0,1)$. However, vulnerability ($E_\sigma''(x)<0$ for $x\to 0,1$)
and/or $E_\sigma''(x)<0$ for all $x\in(0,1)$ is not always present.

For studying the impact of energy landscape features on the activation
energies, we start by analyzing the impact of the mismatch energy
$\epsilon_{\rms mis}$ on the behavior of $E_\alpha^{\rm AHL}(x)$ in
the absence of onsite energy correlations ($\rho=0$) and spatial
correlations ($\kappa=0$).  Corresponding results are shown for
$\epsilon_{\rms mis}=0$ and 3 in Fig.~\ref{fig:model_parameters}(a) in
the absence of both onsite energy correlations ($\rho=0$) and spatial
correlations ($\kappa=0$).

Remarkably, even in the absence of an energy mismatch ($\epsilon_{\rms
  mis}=0$), a mixed alkali effect occurs: a replacement of one type of
mobile ion by the the other type always leads to an increase and
decrease of the activation energy of the former and latter type. The
occurrence of the mixed alkali effect is even more remarkable in view
of the fact that for our symmetric choice of parameters, the energy
landscape or, more precisely, the joint PDF of site energies in
Eq.~\eqref{eq:joint-PDF}, does not change with $x$. This is because
the two bivariate Gaussian distributions $G_{\rm A}(\epsilon_{\rm A},
\epsilon_{\rm B})$ and $G_{\rm B}(\epsilon_{\rm A}, \epsilon_{\rm B})$
are equal.

The mixed alkali effect for $\epsilon_{\rm mis}=0$ originates solely
from the variation of the two Fermi energies $\epsilon_{\rms F}^{\rms
  A}$ and $\epsilon_{\rms F}^{\rms B}$ with $x$, causing the effective
barriers in Eq.~\eqref{eq:uij0alpha} [or Eq.~\eqref{eq:u0ij}] to
change.  As the Fermi energies are determined by the two coupled
Eqs.~\eqref{eq:mu-determ-2-a}, \eqref{eq:mu-determ-2-b}, one can also
argue that the generalized Fermi distribution
\eqref{eq:two-species-fermi-function} is decisive. Said differently,
energetic disorder together with a site exclusion interaction is
sufficient for a mixed alkali effect to occur. The presence of an
energetic mismatch ($\epsilon_{\rm mis}=3$), however, leads to a
significantly stronger mixed alkali effect.

Next, we fix $\epsilon_{\rm mis}=3$ and analyze the impact of the
onsite energy correlations on the behavior of the $E_\alpha^{\rms
  AHL}(x)$ in the absence of spatial correlations. Corresponding
results in Fig.~\ref{fig:model_parameters}(b) for both strong
anticorrelations ($\rho=-0.9$) and strong positive correlations
($\rho=0.9$) show that the character of onsite energy correlations
mainly modifies $E_\alpha''(x)$ of the minority ion. For $\rho=0.9$,
$E_\alpha^{\rms AHL}(x)$ of the minority ion has a pronounced concave
form. Lowering $\rho$, the concavity decreases, see the curve for
$\rho=0$, $\epsilon_{\rm mis}$ in Fig.~\ref{fig:model_parameters}(a).
For anticorrelated onsite energies, $E_\alpha^{\rms AHL}(x)$ of the
minority ion becomes convex.

Finally, we fix both $\epsilon_{\rm mis}=3$ and $\rho=0.9$ and
consider the impact of spatial correlations between site
energies. Results in Fig.~\ref{fig:model_parameters}(c) for
$\kappa=1/3$ and $\kappa=2/3$ corroborate the earlier finding that
vulnerability can emerge due to spatial energy correlations. While for
$\kappa=0$ in Figs.~\ref{fig:model_parameters}(a,b), $E_\alpha^{\rms
  AHL}(x)$ of the majority ion is convex in the limit $x\to0$ (or
$x\to1$), this behavior turns to a concave one for increasing spatial
correlation strength. With the change of convexity of
$E_\alpha$ for the majority ion goes along a weakening of the onsite
correlation effect on the $E_\alpha^{\rms AHL}(x)$ of the minority
ion.

\subsection{Modeling experimental conductivity\newline activation energies}
\label{subsec:modeling-experimental-observations}
In Eq.~\eqref{eq:joint-PDF}, both bivariate Gaussian distributions
$G_\alpha(\epsilon_{\rm A},\epsilon_{\rm B})$
contain five parameters: two specifying the mean values
$\bar\epsilon^\alpha_{\rm A}$, $\bar\epsilon^\alpha_{\rm B}$, and
three specifying the symmetric covariance matrix. While all of them
have their physical significance, as discussed in
Sec.~\ref{sec:pdf-site-energies}, it is desirable to fit experimental
data with a minimal number of parameters to capture essential
features.

Experimental data for total ionic conductivities and their activation
energies $E_\sigma(x)$ alone do not provide much information on
transport properties of the less mobile ion.  Reasonable estimates for
the onsite correlations $\rho_{\rm A}$ and $\rho_{\rm B}$ are then
difficult to obtain.  This is because $E_\sigma(x)$ equals the
activation energy $E_\alpha(x)$ of the more mobile ion
[Eq.~\eqref{eq:Esigma}], on which onsite energy correlations have a
minor influence, as discussed in
Sec.~\ref{subsec:activation-energy-landscape-features}.  Hence, for
modeling $E_\sigma(x)$ without additional information about the
individual $E_\alpha(x)$, we set $\rho_{\rm A}=\rho_{\rm B}=0$.

\begin{figure}[t!]
\centering
\includegraphics[width=\linewidth]{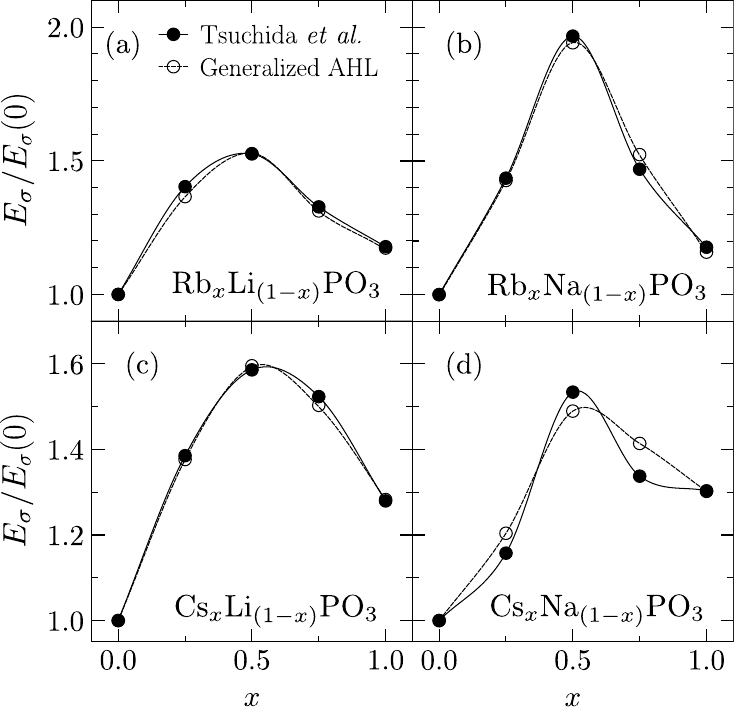}
\caption{\small Fits of total conductivity activation energies
$E_\sigma(x)$ calculated from the generalized AHL theory to
experimental values obtained from impedance measurements reported by
Tsuchida \textit{et al.}~\cite{Tsuchida/etal:2017}.  Energies are
normalized by the activation energies for $x=0$.
The optimal parameter sets of the fitting are listed
in Tab.~\ref{tab:fit_parameters}. Lines are guides to the eye.}
\label{fig:Tsuchida_fit}
\end{figure}

Measured $E_\sigma(x)$ are asymmetric with respect to the mixing of
the mobile ions, i.e.\ they do not satisfy $E_\sigma(x)=E_\sigma(1-x)$
as in our symmetric parametric setting used for the analysis of
salient features. For taking into account asymmetry with fewest
parameters, we choose $\Delta^{\rm A}_{\rm A}$ and $\Delta^{\rm B}_{\rm B}$
to be different and set
\begin{subequations}
\label{eq:parametersimplification}
\begin{align}
\bar\epsilon^{\rm A}_{\rm A}&=\bar\epsilon^{\rm B}_{\rm B}=0\,,
\label{eq:parametersimplification-a}\\
\Delta^\alpha_\beta&=\Delta^\alpha_\alpha\equiv\Delta_\alpha\,,\hspace{1em}\beta\ne\alpha\,,
\label{eq:parametersimplification-b}\\
\epsilon_{\rm mis}^{\rm A}&=\epsilon_{\rm mis}^{\rm B}\equiv\epsilon_{\rm mis}\,.
\label{eq:parametersimplification-c}
\end{align}
\end{subequations}
With this setting, only the four parameters $\Delta_{\rm A}$,
$\Delta_{\rm B}$, $\epsilon_{\rm mis}$, and $\kappa$ remain.  As
discussed in Sec.~\ref{sec:pdf-site-energies}, $\Delta_\alpha$
quantifies how strongly the site energies of A and B ions fluctuate on
a site adapted to an $\alpha$ ion. The mismatch energy $\epsilon_{\rm mis}$
specifies the mean additional energy of an ion of one type
when occupying a site adapted to the other type. The parameter
$\kappa$ gives the strength of spatial correlations between site
energies.

\begin{table}[b!]
\begin{tabular}{ccccc}
\toprule
System & $\kappa$ & $\Delta_{\rm A}/\SI{}{\eV}$ & $\Delta_{\rm B}/\SI{}{\eV}$ & $\epsilon_{\rm mis}/\SI{}{\eV}$\\
\midrule
Rb$_x$Li$_{1-x}$PO$_3$ & 0.84 & 4.12 & 3.13 & 3.36 \\
Rb$_x$Na$_{1-x}$PO$_3$ & 0.30 & 3.49 & 1.26 & 1.34 \\
Cs$_x$Li$_{1-x}$PO$_3$ & 0.63 & 3.31 & 2.21 & 2.67 \\
Cs$_x$Na$_{1-x}$PO$_3$ & 0.07 & 1.61 & 1.23 & 2.48 \\
\bottomrule
\end{tabular}
\caption{\small Parameter sets of best theoretical fits to the
measured activation energies shown in Fig.~\ref{fig:Tsuchida_fit}.}
\label{tab:fit_parameters}
\end{table}

For fitting experimental $E_\sigma(x)$, these four parameters can be
determined as follows. Starting with some fixed $\kappa$,
e.g.\ $\kappa=0$, and requiring $E_\sigma^{\rm AHL}(0)$ and
$E_\sigma^{\rm AHL}(1)$ to match $E_\sigma(0)$ and $E_\sigma(1)$,
uniquely determines $\Delta_{\rm A}$ and $\Delta_{\rm B}$.  This is
because $\epsilon_{\rm mis}$ has no meaning for the single ionic
glasses. Having obtained $\Delta_{\rm A}$ and $\Delta_{\rm B}$,
$\epsilon_{\rm mis}$ is obtained by a least square fit of
$E_\sigma^{\rm AHL}(x)$ to $E_\sigma(x)$.  The residual sum of squared
errors is finally minimized by varying $\kappa$, yielding optimized
parameters.

For nonzero $\kappa$, one needs to take into account that the joint
PDF $\psi(\epsilon_{\rm A}, \epsilon_{\rm B})$ from
Eq.~\eqref{eq:joint-PDF} becomes modified, as discussed in
Sec.~\ref{subsec:spatial-correlations} and
Appendix~\ref{app:PDF-correlated-energies}.  If $x=0$ or $x=1$, the
modified $\psi(\epsilon_{\rm A}, \epsilon_{\rm B})$ remains a single
bivariate Gaussian distribution [see Eq.~\eqref{eq:psiprime}], where
for the parameter setting in
Eqs.~\eqref{eq:parametersimplification-a}-\eqref{eq:parametersimplification-c}
the $\Delta_\alpha$ are rescaled to
\begin{equation}
\Delta'_\alpha= \sqrt{\frac{\kappa^2}{6}+(1-\kappa^2)}\,\Delta_{\alpha}\,.
\end{equation}

As an example, we apply the procedure to fit $E_\sigma(x)$ of mixed
alkali phosphate glasses with composition A$_x$B$_{1-x}$PO$_3$, where
${\rm A}={\rm Rb}$ or Cs, and ${\rm B}={\rm Li}$ or Na.  Conductivity
activation energies of these glasses have been measured by Tsuchida
\textit{et al.} \cite{Tsuchida/etal:2017}. As shown in
Fig.~\ref{fig:Tsuchida_fit}, the fitting by the generalized AHL theory
gives very good agreement with the measurements.

The fitting parameters are listed in Table~\ref{tab:fit_parameters}.
As expected, weak spatial correlations between site energies yield
convex curvatures $E_\sigma''(x)>0$, as for Rb-Na and Cs-Na mixing,
and strong spatial correlations are needed for obtaining concave ones,
as for Rb-Li and Cs-Li mixing.  This finding suggests that phosphate
glasses with mobile Li ions tend to exhibit stronger spatial
correlations.

The necessity to take into account spatial correlations leads to
complex dependencies of $E_\sigma(x)$ on the model parameters, because
they strongly affect percolation thresholds, as given by
Eq.~\eqref{eq:Ealpha}.  This fact reflects the complexity observed in
experiments, where, for example, the strength of the mixed alkali
effect is not directly related to the size difference between the
mobile ion types. For the mixed alkali phosphate glasses in
Fig.~\ref{fig:Tsuchida_fit}, for example, $E_\sigma$ shows a much
stronger mixed alkali effect for Rb-Na than for Rb-Li mixing.

\section{Conclusions}
We have devised a general concept for modeling the MAE based on ion
hopping in disordered site energy landscapes with site exclusion
interaction.  Compared to previous studies, statistical mechanical
features implied by ion mixing have been consistently taken into
account.

We introduced a joint probability density function (PDF) of site
energies $\epsilon_{\rm A}, \epsilon_{\rm B}$ of the two mobile ions A
and B. It gives the concentration of sites where each of the mobile
ions has its own specific energy.

The site exclusion interaction implies that the mean occupation
numbers are given by Fermi statistics, where for the ion mixture the
Fermi distribution takes a generalized form. In the $\epsilon_{\rm A},
\epsilon_{\rm B}$-plane of site energies, the Fermi edges are
represented by lines in the zero-temperature limit $T\to0$. The
hopping dynamics at low $T$ takes place on a set of sites in a narrow
band of order $\kb T$ close to these edges.

Ion conduction properties of the mixture can be characterized by the
Onsager kinetic coefficients, where partial conductivities are given
by a sum of diagonal and nondiagonal elements. The partial
conductivities describe the current of one ion species in response to
an applied electric field. The nondiagonal Onsager coefficients take
into account that a force acting on one type of ion gives rise to a
current of the other type. This is a consequence of the interaction
between the mobile ions.

For an analytical description of the ion transport properties, we
generalized the theory of Ambegaokar, Halperin and
Langer~\cite{Ambegaokar/etal:1971} (AHL theory). This theory was originally
developed for electron conduction in amorphous semiconductors based on
an approach first proposed by Miller and Abrahams
\cite{Miller/Abrahams:1960}.  By our generalization, hopping dynamics
in disordered site energy networks are mapped onto hopping dynamics of
noninteracting ions in disordered barrier networks where all site
energies are equal. This mapping is not exact but yields a good
approximation for calculating transport quantities efficiently.  For a
system with a given joint probability of site energies, partial
conductivities and their activation energies in such barrier networks
can be calculated in a few seconds computing time by employing a Kron
reduction \cite{Bosi/Maass:2022} or the velocity autocorrelation
method \cite{Schroeder:2008}.

In the modeling of this study, we have taken a weighted sum of two
bivariate Gaussian distributions for the joint PDF, one referring to
environments preferentially adapted to sites for the A ions and the
other for environments preferentially adapted to sites for the
B ions.  This choice is motivated by the central limit theorem and
earlier structural-informed modeling of site energy distribution in
alkali phosphate glasses.  The weighting is done
with the mixing ratio
$x=N_{\rm A}/(N_{\rm A}+N_{\rm B})$ of the two mobile ions with
numbers $N_{\rm A}, N_{\rm B}$ and takes into account that the number
of sites with preferentially adapted structural environments to A  and B
ions should be proportional to $x$  and $(1-x)$ in a melt-grown
glass. Hopping rates were chosen according to standard chemical
kinetics.

We have determined the Onsager coefficients for this model by KMC
simulations. The simulations showed that the nondiagonal Onsager
coefficients are much smaller than the diagonal ones and can be
neglected to a first approximation. Partial conductivities and their
activation energies are in good agreement with the AHL theory.

Subsequent analysis of conductivity activation energies based on the
AHL theory showed that a MAE can be present even in the absence of a
mismatch energy between the two mobile ions, i.e.\ when the two
bivariate Gaussian distributions in the weighting are equal, implying
that no preferential environments of sites are assumed to exist for A
and B ions. Considering a mismatch energy, however, yields a stronger
mixed alkali effect.

Onsite energy correlations, given by the Pearson correlation
coefficients of the bivariate Gaussian distributions have a
significant impact on the change of partial conductivities with the
mixing ratio $x$ for the minority ion, while hardly influencing
partial conductivities of the majority ion.  For the minority ion,
they lead to concave (convex) curvatures of $E_\alpha(x)$,
$\alpha={\rm A, B}$, for strong positive (negative) onsite energy
correlation.

Curvatures of $E_\alpha(x)$ for the majority ion are influenced by
spatial correlations between site energies.  In agreement with
previous findings \cite{Peibst/etal:2005,Maass/Peibst:2006}, the
vulnerability effect can be accounted for by spatial correlations,
i.e.\ the stronger than exponential decrease of the mobility of the
majority ion in the dilute foreign alkali region.

Because of the fast calculation of conductivities and activation
energies based on the AHL theory, it is possible to perform a detailed
fitting of model parameters to experimental data.  We demonstrated
such fitting for four different mixed alkali phosphate glasses. Among
others, the fitting suggests that site energies are more strongly
correlated in the phosphate glasses containing Li ions. We propose to
test this prediction by solid-state NMR techniques \cite{Eckert:2018}.
The model parameter $\kappa$ should be related to 
the extent of linkages between 
 network former cations with $n$ covalent bonds ($Q_n$ species), 
which are measurable by 
radio frequency driven recoupling and the double 
quantum coherence approach \cite{Eckert/etal:2005}.
Moreover, recent results for other 
mixed alkali phosphate glasses \cite{Jerroudi/etal:2022, Song/etal:2022, Batista/etal:2026} than
those considered in Ref.~\cite{Tsuchida/etal:2017}
should provide interesting insights on how the fitted model parameters
become modified with glass composition.

We anticipate that a similar fitting of model parameters 
will be successful for other mixed-alkali glasses, such as the widely 
studied mixed-alkali silicate \cite{Terai:1971, Moynihan/etal:1980, Lapp/Shelby:1987, Ali/etal:1995}
and borate glasses \cite{Jain/etal:1984, Mishina/etal:1987, Soppe/etal:1988, Gao:2005, Gao:2006, Imre/etal:2006, Kim/etal:2009}. 
Information 
derived from these fitting parameters regarding site energy mismatch, 
site energy fluctuations, and their spatial correlations can be 
validated against molecular dynamics simulations and checked for
consistency with structural data obtained from EXAFS, Raman spectroscopy, 
and solid-state NMR.

The good agreement of the modeling with experimental data may seem
surprising in view of the fact that the Coulomb interaction between
the mobile ions was not taken into account and that a simple cubic
lattice was used to represent the spatial arrangement of sites. As
discussed in the Introduction, the neglect of Coulomb interaction may
be reasoned by its long-range nature and its much weaker contribution
to the spatial variations of site energies compared to the interaction
with the network forming ions.

Details of the spatial arrangement of sites could be important for
short-time / high-frequency transport properties. For long-range
transport properties, the choice of a regular lattice should not be
particularly relevant, because the conduction takes place on
percolation pathways with a highly irregular, fractal-like topology.

The choice of Gaussian distributions in our modeling is reasonable, 
but not necessary for calculating partial conductivities and their
activation energies based on the
generalized AHL theory presented in Sec.~\ref{subsec:generalized-AHL-theory}.
This theory can be applied for any joint site energy distribution and
also any other model of saddle point energies. The general conclusions drawn 
in Sec.~\ref{subsec:activation-energy-landscape-features}
on the behavior of activation energies in dependence of the energetic mismatch and 
spatial correlations 
between site energies should remain valid for other distributions of site energies
 and saddle point energies.

With the increasing computational power, it may
be possible in the future to perform DFT calculations for a large set
of structural environments with different spatial arrangements of
network forming units surrounding a site. Such investigations can
provide important information on microscopic structural properties and site energies.
Transition path analyses may allow for determining saddle point energies
for individual ion jumps.

We have focused in this study on long-range partial conductivities and
their activation energies.  Another quantity showing a striking mixed
alkali effect is the internal friction \cite{Shelby:1965,
  Shelby/Day:1969, Shelby/Day:1970}. The internal friction can be
obtained by calculating time-dependent correlation functions of
occupation numbers \cite{Peibst/etal:2005, Maass/Peibst:2006}.

Frequency-dependent conductivities are quickly calculated in the
generalized AHL theory by applying the velocity autocorrelation
method.  In KMC simulations they can be determined from
current-current correlations in equilibrium or by analyzing ionic
currents in the presence of an oscillating electric field. Internal
friction and frequency-dependent conductivities need to be revisited
based on our modeling concept.

A further perspective is an extension of the approach to glasses with
more than two mobile ions.  This extension can be done by a
straightforward generalization of the methodology for the joint PDF of
sites energies, the Fermi distributions, and the AHL theory.

\begin{acknowledgments}
We gratefully acknowledge financial support by the Deutsche
Forschungsgemeinschaft (Project No.\ 428906592), and the use of a
high-performance computing cluster funded by the Deutsche
Forschungsgemeinschaft (Project No.\ 456666331).
\end{acknowledgments}

\appendix

\section{Fermi distributions for multicomponent systems}
\label{app:fermi-distributions}
For the Hamiltonian in \eqref{eq:H2}, the grand canonical partition
function $\mathcal{Z}$ can be calculated exactly. Defining by
$\mathcal{S}=\{(0,0),(1,0),(0,1)\}$ the set of possible pairs of
occupation numbers $(n_i^{\rm A},n_i^{\rm B})$ for a site $i$, we
obtain
\begin{align}
\mathcal{Z} & =\sum_\mathbf{n}e^{-\beta H(\mathbf{n})}\\[-1ex]
& =\hspace{-1em}\sum_{(n_1^{\rm A},n_1^{\rm B})\in\mathcal{S}}\hspace{0.3em}
\sum_{(n_2^{\rm A},n_2^{\rm B})\in\mathcal{S}}\ldots \sum_{(n_M^{\rm A},n_M^{\rm B})\in\mathcal{S}}\hspace{-3ex}
e^{-\beta \sum\limits_{i=1}^M (\tilde\epsilon_i^{\rm A} n_i^{\rm A}+\tilde\epsilon_i^{\rm B} n_i^{\rm B})}
\nonumber\\[1ex]
& =\prod_{i=1}^M\sum_{(n_i^{\rm A},n_i^{\rm B})\in\mathcal{S}}\hspace{-3ex}
e^{-\beta(\tilde\epsilon_i^{\rm A} n_i^{\rm A}+\tilde\epsilon_i^{\rm B} n_i^{\rm B})}
=\prod_{i=1}^M\hspace{0.5em} (1\!+\!e^{-\beta\tilde\epsilon_i^{\rm A}}\!+\!e^{-\beta\tilde\epsilon_i^{\rm B}})\,.
\nonumber
\end{align}
The grand canonical potential is
\begin{align}
\Omega(T,\mu_{\rm A},\mu_{\rm B},M)&=-\kb T\ln\mathcal{Z}(T,\mu_{\rm A},\mu_{\rm B},M)\\
&=-\kb T\sum_{i=1}^M\ln (1+e^{-\beta\tilde\epsilon_i^{\rm A}}+e^{-\beta\tilde\epsilon_i^{\rm B}})\,,
\nonumber
\end{align}
from which we obtain
\begin{equation}
\langle n_i^{\rm A}\rangle_{\rm eq}
=\frac{\partial\Omega}{\partial\tilde\epsilon_i^{\rm A}}
=f(\beta\tilde\epsilon_i^{\rm A},\beta\tilde\epsilon_i^{\rm B})\,,
\label{eq:meanA-two-species-app}
\end{equation}
with $f(x,y)$ given in Eq.~\eqref{eq:two-species-fermi-function}.

\section{Behavior of chemical potentials at low temperatures}
\label{app:low-T-behavior-mu}
In the limit $T\to0$ ($\beta\to\infty$), we can write for
$f(\beta\tilde\epsilon_{\rm A},\beta\tilde\epsilon_{\rm B})$ in
Eq.~\eqref{eq:meanA-two-species}
\begin{align}
\lim_{\beta\to\infty}f(\beta\tilde\epsilon_{\rm A},\beta\tilde\epsilon_{\rm B})
&=\theta(-\tilde\epsilon_{\rm A})\theta(\tilde\epsilon_{\rm B}\!-\!\tilde\epsilon_{\rm A})\,
\label{eq:fbetainf}
\end{align}
where $\theta$ is the Heaviside step function, i.e.\ $\theta(x)=1$ for
$x\ge0$ and zero otherwise.  For finite $T$, we define the difference
between $f(\beta\tilde\epsilon_{\rm A},\beta\tilde\epsilon_{\rm B})$
and its $T\to0$ limit by $h(\beta\tilde\epsilon_{\rm A},\beta\tilde\epsilon_{\rm B})$,
where
\begin{equation}
h(x,y)=f(x,y)-\theta(-x)\theta(y-x)\,.
\end{equation}

Generalizing Sommerfeld's expansion for the one-component case
\cite{Schwabl:2010}, we write Eqs.\eqref{eq:mu-determ-2-a},
\eqref{eq:mu-determ-2-b} as
\vspace{-0.5ex}\begin{subequations}
\label{eq:mu-det}
\begin{align}
c_{\rm A}=&
\int\limits_{-\infty}^{\mu_{\rm A}}\hspace{-0.2em}\dd\epsilon_{\rm A}\hspace{-1em}
\int\limits_{\epsilon_{\rm A}-\mu_{\rm A}+\mu_{\rm B}}^\infty\hspace{-1.5em}\dd\epsilon_{\rm B}\,
\psi(\epsilon_{\rm A},\epsilon_{\rm B})
\label{eq:mudet-a}\\
&{}+\int\hspace{-0.2em}\dd\epsilon_{\rm A}\hspace{-0.2em}\int\hspace{-0.2em}\dd\epsilon_{\rm B}\,
\psi(\epsilon_{\rm A},\epsilon_{\rm B})\,h(\beta\tilde\epsilon_{\rm A},\beta\tilde\epsilon_{\rm B})\,,
\nonumber\\
c_{\rm B}=&\int\limits_{-\infty}^{\mu_{\rm B}}\hspace{-0.2em}\dd\epsilon_{\rm B}\hspace{-1em}
\int\limits_{\epsilon_{\rm B}-\mu_{\rm B}+\mu_{\rm A}}^\infty\hspace{-1.5em}\dd\epsilon_{\rm A}\,
\psi(\epsilon_{\rm A},\epsilon_{\rm B})
\label{eq:mudet-b}\\
&{}+\int\hspace{-0.2em}\dd\epsilon_{\rm A}\hspace{-0.2em}\int\hspace{-0.2em}\dd\epsilon_{\rm B}\,
\psi(\epsilon_{\rm A},\epsilon_{\rm B})\,h(\beta\tilde\epsilon_{\rm B},\beta\tilde\epsilon_{\rm A})\,.
\nonumber
\end{align}
\end{subequations}
With the substitutions $x=\beta\tilde\epsilon_{\rm A}$, $\epsilon_{\rm A}=\mu_{\rm A}+\kb Tx$, 
and $y=\beta\tilde\epsilon_{\rm B}$, $\epsilon_{\rm B}=\mu_{\rm B}+\kb Ty$, 
the second term on the right-hand side of Eq.~\eqref{eq:mudet-a} becomes
\begin{align}
&\int\hspace{-0.2em}\dd\epsilon_{\rm A}\hspace{-0.2em}\int\hspace{-0.2em}\dd\epsilon_{\rm B}\,
\psi(\epsilon_{\rm A},\epsilon_{\rm B})\,h(\beta\tilde\epsilon_{\rm A},\beta\tilde\epsilon_{\rm B})
\label{eq:secondterm-result}\\
&\hspace{1.5em}{}=(\kb T)^2\hspace{-0.2em}\int\hspace{-0.3em}\dd x\hspace{-0.3em} \int\hspace{-0.3em}\dd y\,
\psi(\mu_{\rm A}\!+\!\kb Tx,\mu_{\rm B}\!+\!\kb Ty)h(x,y)
\nonumber\\
&\hspace{1.5em}{}\eqo\; (\kb T)^2 \psi(\epsilon_{\rms F}^{\rm A},\epsilon_{\rms F}^{\rm B})
\hspace{-0.2em}\int\hspace{-0.3em}\dd x\hspace{-0.3em} \int\hspace{-0.3em}\dd y\,h(x,y)
=-(\kb T)^2 \zeta_0\,\psi_{\rms F}\,,
\nonumber
\end{align}
where $\psi_{\rms F}=\psi(\epsilon_{\rms F}^{\rm A},\epsilon_{\rms F}^{\rm B})$ and
the relation ``$\eqo$'' here means equal in leading
order with respect to $T$.
The constant $\zeta_0$ is
\begin{equation}
\zeta_0=-\hspace{-0.2em}\int\hspace{-0.3em}\dd x\hspace{-0.3em} \int\hspace{-0.3em}\dd y\,
h(x,y)\cong0.822\,.
\label{eq:kappa}
\end{equation}
The second term on the right-hand side of Eq.~\eqref{eq:mudet-b} gives
the same leading order correction~\eqref{eq:secondterm-result}
because in a derivation analogous to the steps in
Eq.~\eqref{eq:secondterm-result} only $x$ and $y$ in the expression
$h(x,y)$ become interchanged, which does not alter $\zeta_0$.

To tackle the first terms on the right-hand sides of
Eqs.~\eqref{eq:mudet-a}, \eqref{eq:mudet-b}, we define
\vspace{-1ex}\begin{subequations}
\label{eq:Phi-definition}
\begin{align}
\Phi_{\rm A}(\mu_{\rm A},\mu_{\rm B})&=
\int\limits_{-\infty}^{\mu_{\rm A}}\hspace{-0.2em}\dd\epsilon_{\rm A}\hspace{-1em}
\int\limits_{\epsilon_{\rm A}-\mu_{\rm A}+\mu_{\rm B}}^\infty\hspace{-1.5em}\dd\epsilon_{\rm B}\,
\psi(\epsilon_{\rm A},\epsilon_{\rm B})\label{eq:GA-definition}\\
&=\int\limits_{-\infty}^0\hspace{-0.2em}\dd\epsilon_{\rm A}\hspace{-0.2em}
\int\limits_{\epsilon_{\rm A}}^\infty\hspace{-0.3em}\dd\epsilon_{\rm B}\,
\psi(\epsilon_{\rm A}+\mu_{\rm A},\epsilon_{\rm B}+\mu_{\rm B})\,,
\nonumber\\
\Phi_{\rm B}(\mu_{\rm A},\mu_{\rm B})&=
\int\limits_{-\infty}^{\mu_{\rm B}}\hspace{-0.2em}\dd\epsilon_{\rm B}\hspace{-1em}
\int\limits_{\epsilon_{\rm B}-\mu_{\rm B}+\mu_{\rm A}}^\infty\hspace{-1.5em}\dd\epsilon_{\rm B}\,
\psi(\epsilon_{\rm A},\epsilon_{\rm B})\label{eq:GB-definition}\\
&=\int\limits_{-\infty}^0\hspace{-0.2em}\dd\epsilon_{\rm B}\hspace{-0.2em}
\int\limits_{\epsilon_{\rm B}}^\infty\hspace{-0.3em}\dd\epsilon_{\rm A}\,
\psi(\epsilon_{\rm A}+\mu_{\rm A},\epsilon_{\rm B}+\mu_{\rm B})\,,
\nonumber
\end{align}
\end{subequations}
and Taylor-expand these functions around 
$(\epsilon_{\rms F}^{\rm A},\epsilon_{\rms F}^{\rm B})$ up to first order, taking into
account $\Phi_{\rm A}(\epsilon_{\rms F}^{\rm A},\epsilon_{\rms F}^{\rm
  B})=c_{\rm A}$ and 
  $\Phi_{\rm B}(\epsilon_{\rms F}^A,\epsilon_{\rms F}^B)=c_{\rm B}$:
\begin{subequations}
\label{eq:G-expansion}
\begin{align}
\Phi_{\rm A}(\mu_{\rm A},\mu_{\rm B})&\eqo c_{\rm A}
+\varphi_{\rm AA}(\mu_{\rm A}-\epsilon_{\rms F}^{\rm A})
+\varphi_{\rm AB}(\mu_{\rm B}-\epsilon_{\rms F}^{\rm B})\,,
\label{eq:GA-expansion}\\[0.5ex]
\Phi_{\rm B}(\mu_{\rm A},\mu_{\rm B})&\eqo c_{\rm B}
+\varphi_{\rm BA}(\mu_{\rm A}-\epsilon_{\rms F}^{\rm A})
+\varphi_{\rm BB}(\mu_{\rm B}-\epsilon_{\rms F}^{\rm B})\,.
\label{eq:GB-expansion}
\end{align}
\end{subequations}

\vspace{-1ex}\noindent
Here,
\begin{subequations}
\label{eq:g-coefficients}
\begin{align}
\varphi_{{\rm A},\alpha}&
=\varphi_{{\rm A},\alpha}(\epsilon_{\rms F}^{\rm A},\epsilon_{\rms F}^{\rm B})
=\partial_{\mu_\alpha} \Phi_{\rm A}(\mu_{\rm A},\mu_{\rm B})
\Bigr|_{\mbox{$\sss (\mu_{\rm A},\mu_{\rm B})=(\epsilon_{\rms F}^{\rm A},\epsilon_{\rms F}^{\rm B})$}}
\nonumber\\
&=\int\limits_{-\infty}^0\hspace{-0.2em}\dd\epsilon_{\rm A}\hspace{-0.2em}
\int\limits_{\epsilon_{\rm A}}^\infty\hspace{-0.3em}\dd\epsilon_{\rm B}\,
\partial_{\epsilon_\alpha}\psi(\epsilon_{\rm A}+\epsilon_{\rms F}^{\rm A},\epsilon_{\rm B}+\epsilon_{\rms F}^{\rm B})\,,
\label{eq:gA-coefficients}\\[1ex]
\varphi_{{\rm B},\alpha}&=\varphi_{{\rm B},\alpha}(\epsilon_{\rms F}^{\rm A},\epsilon_{\rms F}^{\rm B})
=\partial_{\mu_\alpha} \Phi_{\rm B}(\mu_{\rm A},\mu_{\rm B})
\Bigr|_{\mbox{$\sss (\mu_{\rm A},\mu_{\rm B})=(\epsilon_{\rms F}^{\rm A},\epsilon_{\rms F}^{\rm B})$}}
\nonumber\\
&=\int\limits_{-\infty}^0\hspace{-0.2em}\dd\epsilon_{\rm B}\hspace{-0.2em}
\int\limits_{\epsilon_{\rm B}}^\infty\hspace{-0.3em}\dd\epsilon_{\rm A}\,
\partial_{\epsilon_\alpha}\psi(\epsilon_{\rm A}
+\epsilon_{\rms F}^{\rm A},\epsilon_{\rm B}+\epsilon_{\rms F}^{\rm B})\,.
\label{eq:gB-coefficients}
\end{align}
\end{subequations}

By inserting Eqs.~\eqref{eq:secondterm-result}, \eqref{eq:G-expansion}
into Eq.~\eqref{eq:mudet-a}, \eqref{eq:mudet-b} we obtain
\vspace{-1ex}\begin{subequations}
\begin{align}
\varphi_{\rm AA}(\mu_{\rm A}-\epsilon_{\rms F}^{\rm A})+\varphi_{\rm AB}(\mu_{\rm B}-\epsilon_{\rms F}^{\rm B})&
=(\kb T)^2\zeta_0\,\psi_{\rms F}\,,\\
\varphi_{\rm BA}(\mu_{\rm A}-\epsilon_{\rms F}^{\rm A})+\varphi_{\rm BB}(\mu_{\rm B}-\epsilon_{\rms F}^{\rm B})&
=(\kb T)^2\zeta_0\,\psi_{\rms F}\,.
\end{align} 
\end{subequations}
By solving these linear equations for the chemical potentials, we obtain
\vspace{-1ex}\begin{subequations}
\begin{align}
\mu_{\rm A}&\,\eqo\,\epsilon_{\rms F}^{\rm A}+(\kb T)^2\zeta_0\,\psi_{\rms F}\,
\frac{\mdet{
1 & \varphi_{\rm AB}\\
1 & \varphi_{\rm BB}}
}
{\mdet{
\varphi_{\rm AA} & \varphi_{\rm AB}\\
\varphi_{\rm BA} & \varphi_{\rm BB}}
}\nonumber\\
&=\epsilon_{\rms F}^{\rm A}+(\kb T)^2\zeta_0\,\psi_{\rms F}\,\frac{\varphi_{\rm BB}
-\varphi_{\rm AB}}{\varphi_{\rm AA}\varphi_{\rm BB}-\varphi_{\rm AB}\varphi_{\rm BA}}\,,
\\[0.5ex]
\mu_{\rm B}&\,\eqo\,\epsilon_{\rms F}^{\rm B}+(\kb T)^2\zeta_0\,\psi_{\rms F}\,
\frac{\mdet{
\varphi_{\rm AA}  & 1\\
\varphi_{\rm BA} & 1}
}
{\mdet{
\varphi_{\rm AA} & \varphi_{\rm AB}\\
\varphi_{\rm BA} & \varphi_{\rm BB}}
}\nonumber\\
&=\epsilon_{\rms F}^{\rm B}+(\kb T)^2\zeta_0\,\psi_{\rms F}\,\frac{\varphi_{\rm AA}
-\varphi_{\rm BA}}{\varphi_{\rm AA}\varphi_{\rm BB}-\varphi_{\rm AB}\varphi_{\rm BA}}\,.
\end{align}
\end{subequations}
This yields Eq.~\eqref{eq:mu-alpha-low-T} with
\begin{subequations}
\begin{align}
\zeta_{\rm A}&=\zeta_0\,\psi_{\rms F}\,\frac{\varphi_{\rm BB}
-\varphi_{\rm AB}}{\varphi_{\rm AA}\varphi_{\rm BB}-\varphi_{\rm AB}\varphi_{\rm BA}}\,,\\
\zeta_{\rm B}&=\zeta_0\,\psi_{\rms F}\,\frac{\varphi_{\rm AA}
-\varphi_{\rm BA}}{\varphi_{\rm AA}\varphi_{\rm BB}-\varphi_{\rm AB}\varphi_{\rm BA}}\,.
\end{align} 
\end{subequations}

\section{Joint probability density of spatially correlated site energies}
\label{app:PDF-correlated-energies}
When combining the two site energies $\epsilon_i^{\rm A}$ and
$\epsilon_i^{\rm B}$ into pairs $(\epsilon_i^{\rm A}, \epsilon_i^{\rm
  B})$, these are uncorrelated pairs of random variables with PDF
$\psi(\epsA,\epsB)$ given in Eq.~\eqref{eq:Galpha}.  The locally
averaged site energies $(\epsAp,\epsBp)$ in
Eq.~\eqref{eq:epsprime-eps} are obtained by summation over the
uncorrelated pairs $(\epsA,\epsB)$.  The PDF $\psi'(\epsAp,\epsBp)$ of
the $(\epsAp, \epsBp)$ can thus be obtained by calculating its
characteristic function, which is a product of scaled characteristic
functions of $\psi(\epsA,\epsB)$.

The characteristic function $\langle e^{i(\omA\epsA+\omB\epsB)}\rangle$
of $\psi(\epsA,\epsB)$ is
\begin{equation}
\hat\psi(\omA,\omB)=x\,\hat G_{\rm A}(\omA,\omB)+(1\!-\!x)\hat G_{\rm B}(\omA,\omB)\,,
\label{eq:characteristic-of-psi}
\end{equation}
where
\begin{equation}
\hat G_\gamma(\omA,\omB)=\exp[i(\omA\bar\epsilon^\gamma_{\rm A}+\omB\bar\epsilon^\gamma_{\rm B})
-\frac{1}{2}\hspace{-0.5em}\sum_{\alpha,\beta={\rm A,B}}
\hspace{-0.5em}\omega_\alpha C^\gamma_{\alpha\beta}\omega_\beta]
\end{equation}
are the characteristic functions of the bivariate Gaussians in
Eq.~\eqref{eq:Galpha}.

For convenient notation, we introduce
\begin{equation}
\kappa_0=1-\kappa\,,\hspace{1em}\kappa_1=\frac{\kappa}{6}\,.
\end{equation}
The characteristic function 
of the PDF $\psi'(\epsAp,\epsBp)$ can then be written as
\begin{equation}
\hat\psi'(\omA,\omB)=\hat\psi(\kappa_0\omA,\kappa_0\omB)\hat\psi(\kappa_1\omA,\kappa_1\omB)^6\,.
\end{equation}
Inserting $\hat\psi(\omA,\omB)$ from
Eq.~\eqref{eq:characteristic-of-psi}, factorizing the term $[x\,\hat
  G_{\rm A}(\kappa_1\omA,\kappa_1\omB)+(1\!-\!x)\hat G_{\rm
    B}(\kappa_1\omA,\kappa_1\omB)]^6$ out, and using
\begin{align}
\hat G_\gamma(\kappa\omA,\kappa\omB)^j&=
\exp\Biggl[ij\kappa(\omA\bar\epsilon^\gamma_{\rm A}+\omB\bar\epsilon^\gamma_{\rm B})\nonumber\\
&\hspace{3em}{}-\frac{j\kappa^2}{2}\hspace{-0.5em}\sum_{\alpha,\beta={\rm A,B}}
\hspace{-0.5em}\omega_\alpha C^\gamma_{\alpha\beta}\omega_\beta\Biggr],
\label{eq:hatG^j}
\end{align}
we obtain
\begin{align}
\hat\psi'(\omA,\omB)&=
\sum_{j=0}^6\mybinom{6}{j}x^j(1\!-\!x)^{6-j}
\label{eq:hatpsiprime}\\
&\hspace{1em}
{}\times\left[x\,\hat  G_{\rm I}^{(j)}(\omA,\omB)+(1\!-\!x)\hat G_{\rm II}^{(j)}(\omA,\omB)\right]\,,
\nonumber
\end{align}
where
\begin{align}
&\hat G_{\rm I}^{(j)}(\omA,\omB)=\label{eq:hatGI}\\
&\hspace{2em}\exp\Biggl\{i\Bigl(\left[(\kappa_0\!+\!j\kappa_1)\mAA+(6\!-\!j)\kappa_1\mBA\right]\omA
\nonumber\\
&\hspace{4em}{}+\left[(\kappa_0\!+\!j\kappa_1)\mAB+(6\!-\!j)\kappa_1\mBB\right]\omB\Bigr)
\nonumber\\
&\hspace{3em}{}
-\frac{1}{2}\hspace{-0.5em}\sum_{\alpha,\beta={\rm A,B}}\hspace{-0.5em}
\omega_\alpha\!\left[(\kappa_0^2\!+\!j\kappa_1^2)C^{\rm A}_{\alpha\beta}+(6\!-\!j)\kappa_1^2C^{\rm B}_{\alpha\beta}\right]\!\omega_\beta\Biggr\}\,,
\nonumber\\[2ex]
&\hat G_{\rm II}^{(j)}(\omA,\omB)=\label{eq:hatGII}\\
&\hspace{2em}
\exp\Biggl\{i\Bigl(\left[j\kappa_1\mAA+(\kappa_0\!+\!(6\!-\!j)\kappa_1)\mBA\right]\omA
\nonumber\\
&\hspace{4em}{}+\left[j\kappa_1\mAB+(\kappa_0\!+\!(6\!-\!j)\kappa_1)\mBB\right]\omB\Bigr)
\nonumber\\
&\hspace{3em}{}
-\frac{1}{2}\hspace{-0.5em}\sum_{\alpha,\beta={\rm A,B}}\hspace{-0.5em}
\omega_\alpha \left[j\kappa_1^2C^{\rm A}_{\alpha\beta}+(\kappa_0^2\!+\!(6\!-\!j)\kappa_1^2)C^{\rm B}_{\alpha\beta}\right]\omega_\beta\Biggr\}\,.
\nonumber
\end{align}
Like $\hat\psi(\omA,\omB)$, the characteristic function
$\hat\psi'(\omA,\omB)$ in Eq.~\eqref{eq:hatpsiprime} is a sum of
characteristic functions of bivariate Gaussians.  Hence,
\begin{align}
\psi'(\epsAp,\epsBp)&=\sum_{j=0}^6\mybinom{6}{j}x^j(1\!-\!x)^{6-j}
\label{eq:psiprime}\\
&\hspace{2em}{}\times\left[x\,G_{\rm I}^{(j)}(\epsAp,\epsBp)+(1\!-\!x)G_{\rm II}^{(j)}(\epsAp,\epsBp)\right]\,,
\nonumber
\end{align}
where $G_{\rm I}^{(j)}(\epsAp,\epsBp)$ and $G_{\rm II}^{(j)}(\epsAp,\epsBp)$
are bivariate Gaussians with mean values
\vspace{-1ex}\begin{subequations}
\label{eq:epsI-epsII}
\begin{align}
\bar\epsilon^{\,{\rm I}(j)}_{\rm A}&=(\kappa_0\!+\!j\kappa_1)\mAA+(6\!-\!j)\kappa_1\mBA\,,\\
\bar\epsilon^{\,{\rm I}(j)}_{\rm B}&=(\kappa_0\!+\!j\kappa_1)\mAB+(6\!-\!j)\kappa_1\mBB\,,\\[1ex]
\bar\epsilon^{\,{\rm II}(j)}_{\rm A}&=j\kappa_1\mAA+[\kappa_0\!+\!(6\!-\!j)\kappa_1]\mBA\,,\\
\bar\epsilon^{\,{\rm II}(j)}_{\rm B}&=j\kappa_1\mAB+[\kappa_0\!+\!(6\!-\!j)\kappa_1]\mBB\,,
\end{align}
\end{subequations}
and covariance matrices
\vspace{-1ex}\begin{subequations}
\label{eq:CI-CII}
\begin{align}
C^{{\rm I}(j)}&=(\kappa_0^2\!+\!j\kappa_1^2)\,C^{\rm A}+(6\!-\!j)\kappa_1^2\,C^{\rm B}\,,\\
C^{{\rm II}(j)}&=j\kappa_1^2\,C^{\rm A}+\left[\kappa_0^2\!+\!(6\!-\!j)\kappa_1^2\right]C^{\rm B}\,.
\end{align}
\end{subequations}

\vspace{2ex}
\section{Derivation of Eq.~\eqref{eq:Iijalpha} based on fluctuation-disspation theorem}
\label{app:FDT-derivation_Iij}
It is instructive to see that the derivation of
Eq.~\eqref{eq:ec-potential-difference} can be done alternatively by
connecting $\delta\langle n_i^\alpha\rangle =\langle
n_i^\alpha\rangle-\langle n_i^\alpha\rangle_{\rm eq}$ with the
$\delta\mu_i^\alpha$ via the fluctuation-dissipation theorem (FDT).
In the linear response limit, it holds for $\gamma\ne\alpha$
\begin{equation}
\delta \langle n_i^\alpha\rangle=\langle \delta n_i^\alpha\rangle=
\chi_{\alpha\alpha}\delta\mu_i^\alpha+\chi_{\alpha\gamma}\delta\mu_i^\gamma\,,
\label{eq:deltan-deltamu}
\end{equation}
where
\begin{subequations}
\label{eq:chi}
\begin{gather}
\chi_{\alpha\alpha}=\beta\langle\delta n_i^\alpha\delta n_i^\alpha\rangle_{\rm eq}
=\beta\langle n_i^\alpha\rangle_{\rm eq}(1\!-\!\langle n_i^\alpha\rangle_{\rm eq})\,,
\label{eq:chialphaalpha}\\[0.5ex]
\chi_{\alpha\gamma}=\beta\langle\delta n_i^\alpha\delta n_i^\gamma\rangle_{\rm eq}=
-\beta\langle n_i^\alpha\rangle_{\rm eq} \langle n_i^\gamma\rangle_{\rm eq}
=\chi_{\gamma\alpha}\,.
\label{eq:chialphagamma}
\end{gather}
\end{subequations}
Combining Eqs.~\eqref{eq:deltan-deltamu} and \eqref{eq:chi} yields
\begin{equation}
\delta \langle n_i^\alpha\rangle=\langle n_i^\alpha\rangle_{\rm eq}(1\!-\!\langle n_i^\alpha\rangle_{\rm eq})\,\beta \delta\mu_i^\alpha
-\langle n_i^\alpha\rangle_{\rm eq} \langle n_i^\gamma\rangle_{\rm eq}\,\beta\delta\mu_i^\gamma\,.
\label{eq:dni-FDT}
\end{equation}
The directed link current $I_{i\to j}^\alpha$ from
Eq.~\eqref{eq:Iij-directed} can be written as ($\alpha\ne\gamma$)
\begin{align}
I_{i\to j}^\alpha&=w_{ij}^\alpha\langle n_i^\alpha (1\!-\!n_j^\alpha\!-\!n_j^\gamma)\rangle
\simeq w_{ij}^\alpha\langle n_i^\alpha\rangle (1\!-\!\langle n_j^\alpha\rangle\!-\!\langle n_j^\gamma\rangle)\nonumber\\
&\eqo I_{i\to j}^{0,\alpha}\left(1+\frac{\delta w_{ij}^\alpha}{w_{ij}^{0,\alpha}}+\frac{\langle\delta n_i^\alpha\rangle}{\langle n_i^\alpha\rangle_{\rm eq}}
-\frac{\delta\langle n_j^\alpha\rangle+\delta\langle n_j^\gamma\rangle}
        {1\!-\!\langle n_j^\alpha\rangle_{\rm eq}\!-\!\langle n_j^\gamma\rangle_{\rm eq}}\right).
\label{eq:Iitojalpha-app1}
\end{align}
Inserting the $\delta\langle n_i^\alpha\rangle$ from
Eq.~\eqref{eq:dni-FDT} gives
\begin{align}
I_{i\to j}^\alpha\eqo I_{i\to j}^{0,\alpha}\Bigl(&1+\frac{\delta w_{ij}^\alpha}{w_{ij}^{0,\alpha}}+\beta\delta\mu_i^\alpha-\langle n_i^\alpha\rangle_{\rm eq}\,\beta\delta\mu_i^\alpha
\label{eq:Iitojalpha-app2}\\
&\hspace{-2em}{}-\langle n_i^\gamma\rangle_{\rm eq}\,\beta\delta\mu_i^\gamma
-\langle n_j^\alpha\rangle_{\rm eq}\,\beta\delta\mu_j^\alpha-\langle n_j^\gamma\rangle_{\rm eq}\,\beta\delta\mu_j^\gamma\Bigr)
\nonumber
\end{align}

\vspace{-3ex}\noindent
and
\vspace{-1ex}
\begin{equation}
I_{ij}^\alpha=I_{i\to j}^\alpha-I_{j\to i}^\alpha
\eqo
I_{i\to j}^{0,\alpha}\left[\frac{\delta w_{ij}^\alpha}{w_{ij}^{0,\alpha}}-\frac{\delta w_{ji}^\alpha}{w_{ji}^{0,\alpha}}
+\beta(\mu_i^\alpha\!-\!\mu_j^\alpha)\right],
\label{eq:Iijalpha-app}
\end{equation}
in agreement with Eq.~\eqref{eq:Iijalpha}.

\end{document}